\documentclass{article}
\usepackage{amsfonts}
\usepackage{amsmath}
\usepackage{harvard}

\newenvironment{proof}[1][Proof]{\noindent\textbf{#1.} }{\ \rule{0.5em}{0.5em}}
\input{tcilatex}

\begin{document}

\title{\textbf{Influence of the material substructure on crack propagation:
a unified treatment.}}
\author{Paolo Maria Mariano \\
Dipartimento di Ingegneria Strutturale e Geotecnica,\\
Universit\`{a} di Roma "La Sapienza",\\
via Eudossiana 18, I-00184 Roma (Italy);\\
e-mail: paolo.mariano@uniroma1.it}
\maketitle

\begin{abstract}
The influence of the material texture (substructure) on the force driving
the crack tip in complex materials admitting Ginzburg-Landau-like energies
is analyzed in a three-dimensional continuum setting. The theory proposed
accounts for finite deformations and general coarse-grained order
parameters. A modified expression of the \textsf{J}-integral is obtained
together with other path-integrals which are necessary to treat cases where
the process zone around the tip has finite size. The results can be applied
to a wide class of material substructures. As examples, cracks in
ferroelectrics and in materials with strain-gradient effects are discussed:
in these cases the specializations of the general results fit reasonably
experimental data.
\end{abstract}

\section{Introduction}

To analyze the behavior of cracks, one needs to understand how the
interactions in the material cooperate to drive the crack tip. In the common
setting of deformable simple bodies (Cauchy's model) the question has been
clarified in many aspects from theoretical and computational points of view 
\footnote{%
From the pioneer works of Griffith (1920), Atkinson \& Eshelby (1968),
Baremblatt (1972), Freund (1972), Rice (1968), to contributions like
(Adda-Bedia, Arias, Ben Amar \& Lund, 1999; Dolbow, Mo\"{e}s \& Belytschko,
2001; Freund, 1990; Gurtin \& Shvartsman, 1997; Heino \& Kaski, 1997; Mo\"{e}%
s \& Belytschko, 2002; Obrezanova, Movchan \& Willis, 2002; Oleaga, 2003;
Slepyan, 2002).}. It is not so for complex bodies where experiments suggest
that the influence of material texture (substructure) on the expression of
the `force' driving the crack tip may be prominent. For instance, for
materials that fail by decohesion or cleavage at the atomic scale, the
predictions of the standard theory of fracture are not satisfactory 
\footnote{%
See the remarks about the interpretation of experimental results in (Wei \&
Hutchinson, 1997).}, and the same inaccuracy occurs in the case of cracks
propagating along metal-ceramic interfaces. Moreover, e.g., the spontaneous
polarization in ferroelectrics influences rather stronghly the crack growth 
\footnote{%
Experimental data and detailed analyses are collected in (Fulton \& Gao,
2001).}.

The investigation is matter of continuum theories more intricated than the
standard elasticity where only some global aspects of the material texture
are considered through material symmetries, and a direct description of the
configuration of the substructure (together with its changes) and the
related interactions is absent.

However, if one would like to determine the appropriate expression of the
force driving the crack tip in each special case of complex material, one
would have to construct only a zoo of possible modifications of the standard
theory of fracture for small and large strains, obscuring the basic fact
that there exists a common unifying physical mechanism. The aim of the
present paper is just to show such a mechanism common to all complex
materials admitting Ginzburg-Landau-like energies and to provide accordingly
a general expression of the driving force which can be specified and tested
in special cases. We consider large strains to encompass cases in which they
are prominent as, e.g., in presence of elastomers.

In what follows, an order parameter $\mathbf{\nu }$\ is assigned to each
material element as coarse grained geometrical descriptor of the
substructure. To maintain generality, we require only that $\mathbf{\nu }$\
belongs to a differentiable manifold $\mathcal{M}$. Each special choice of $%
\mathcal{M}$ characterizes the model of each complex material. Interactions
developing \emph{extra power} on the rate of the order parameter, i.e., on
the rate of change of the substructure, are considered and satisfy
appropriate balance equations. Models of fiber-reinforced composites,
ferroelectric and magnetoelastic solids, interacting elastomers,
microcracked and multiphase solids fall, e.g., within such a general
approach.

In what follows, first we derive balance equations at the crack tip for the
interactions generated by the substructure, considering possible
substructural inertia effects at the tip. Then we determine the contribution
of substructural interactions to the expression of the driving force at the
tip of the crack. It results in a modified expression of the \textsf{J}%
-integral: 
\begin{equation*}
\mathsf{J}\mathbf{=}\mathsf{n}\cdot \int_{tip}\left( \rho \left( \frac{1}{2}%
\left\Vert \mathbf{\dot{x}}\right\Vert ^{2}+k\left( \mathbf{\nu },\mathbf{%
\dot{\nu}}\right) \right) \mathbf{I+}\mathbb{P}\right) \mathbf{n,}
\end{equation*}%
where $\mathsf{n}$\ is the direction of propagation of the crack, the
second-order tensor $\mathbb{P}$ is given by $\mathbb{P}=$\ $\psi \mathbf{I-F%
}^{T}\mathbf{T-}\nabla \mathbf{\nu }^{T}\underline{\ast }\mathcal{S}$, with $%
\psi $ the free energy density, $\mathbf{F}$ the gradient of deformation, $%
\mathbf{T}$ the first Piola-Kirchhoff stress, $\mathcal{S}$ a measure of
substructural interactions called microstress, $\mathbf{I}$ the unit tensor, 
$\rho $\ the density of mass, while the other terms account for standard and
substructural kinetic energy (if existent). $\int_{tip}$\ indicates a
special limit process which consists in evaluating an integral on the
boundary of a disc centered at the tip in a plane orthogonal to the tangent
of the tip at a given point and in shrinking the disc up to the tip. The
product $\underline{\ast }$\ is defined below. Special expressions for the 
\textsf{J}-integral follow once the order parameter is specified. Here, to
suggest examples, ferroelectrics and materials with strain-gradient effects
are considered: the relevant \textsf{J}-integrals provide values of the
driving force close to experimental data. Other path-integrals besides 
\textsf{J}-integral are obtained: they allow us to express the energy
dissipated during the evolution of the crack in the case in which the
process zone around the tip has finite size. In the expression of \textsf{J}%
, two elements mark the difference with the standard theory of fracture,
namely the densities $\rho k\left( \mathbf{\nu },\mathbf{\dot{\nu}}\right) $
and $\nabla \mathbf{\nu }^{T}\underline{\ast }\mathcal{S}$ that underline
the explicit influence of the material substructure. Whilst the former is
negligible unless the substructure oscillates at very high frequencies, the
latter may be crucial to justify the discrepancies between experimental data
and the previsions of the standard fracture mechanics. (To render the
formulas as concise as possible, in the integrals we do not write explicitly
line, area and volume differentials: the kind of integration is clear
looking at the domain of integration directly\footnote{\emph{Notations}.
Some standard notations are summarized here. Non-standard notations are
introduced in detail in the rest of the paper. Let $\mathbf{A}$ and $\mathbf{%
B}$ be tensors of type $\left( p,q\right) $ of components e.g. $%
A_{j_{1}...j_{q}}^{i_{1}...i_{p}}$. We denote with $\mathbf{A\cdot B}$ the
standard scalar product given by $%
A_{j_{1}...j_{q}}^{i_{1}...i_{p}}B_{j_{1}...j_{q}}^{i_{1}...i_{p}}$. In
particular, if $\mathbf{A}$ and $\mathbf{B}$ are second order tensors, we
denote with $\mathbf{AB}$ the product which contracts only one index and
bears a second order tensor; for example, we have $\left( \mathbf{AB}\right)
_{ij}=A_{ik}B_{kj}$. If $\mathbf{A}$ and $\mathbf{B}$ are third order
tensors, we indicate with $\mathbf{A:B}$ the product contracting two indices
and bearing a second order tensor. If $\mathbf{A}$ is a tensor of the type $%
\left( p,q\right) $, with $p,q>0$, and $\mathbf{B}$ is another tensor of the
type $\left( r,s\right) $, with $r,s>0$ and $r<p$, $s<q$, or $\left(
r=p,s<q\right) $ or $\left( r<p,s=q\right) $, we indicate with $\mathbf{AB}$
(with some slight abuse of notation with respect to the product between
second order tensors) the product which contracts all the indices of $%
\mathbf{B}$; in particular, if $p=0$ or $q=0$ we take $r=0$ and $s=0$
respectively. Given two vectors $\mathbf{a}$ and $\mathbf{b}$, $\mathbf{%
a\otimes b}$ denotes their tensor product. In particular, if $\mathbf{A}$
and $\mathbf{B}$ are second order tensors we have $\mathbf{AB\cdot }\left( 
\mathbf{a\otimes b}\right) =\mathbf{A}^{T}\mathbf{a\cdot Bb}$. For any
region $\mathfrak{b}$ of the space, $\partial \mathfrak{b}$ represents its
boundary.}.)

\section{Order parameters for the substructure and the geometry of the crack}

Let $\mathcal{B}$\ be the regular region of the three-dimensional Euclidean
point space $\mathcal{E}^{3}$ occupied by the body in its reference place. A
generic point of $\mathcal{B}$ is indicated with $\mathbf{X}$. A `standard'
deformation is described by a sufficiently smooth bijection $\mathcal{B}\ni 
\mathbf{X}\longmapsto \mathbf{x}\left( \mathbf{X}\right) \in \mathcal{E}^{3}$
(with the current place $\mathbf{x}\left( \mathcal{B}\right) $\ a regular
region) which is also orientation-preserving in the sense that the gradient $%
\nabla \mathbf{x}$ of $\mathbf{x}$ with respect to $\mathbf{X}$ (indicated
with $\mathbf{F}$) is such that $\det \mathbf{F}>0$.

As anticipated above, information on the substructure of each material
element are given through the assignment of an \emph{order parameter} $%
\mathbf{\nu }$, by means of a sufficiently smooth mapping $\mathcal{B}\ni 
\mathbf{X}\longmapsto \mathbf{\nu \left( X\right) }\in \mathcal{M}$, where $%
\mathcal{M}$ is a finite dimensional differentiable paracompact manifold
without boundary endowed with metric and connection (Capriz, 1989). The
choice of $\mathcal{M}$\ determines the characteristic features of each
special model of substructure. Vector order parameters with unit length may
represent stiff microfibers in composites with a softer matrix or
magnetostrictive materials. Vector order parameters not constrained to have
unit length are used for ferroelectrics, elastic microcracked bodies,
nematic elastomers. Second order tensor valued order parameters may also
serve as descriptors of families of polymeric chains or polymer stars.

Let $\mathbf{\dot{x}}\left( \mathbf{X,}t\right) $ and $\mathbf{\dot{\nu}}%
\left( \mathbf{X,}t\right) $\ be the velocity and the rate of the order
parameter (in their referential description) evaluated by a given observer.
After a change in observers ruled by $SO\left( 3\right) $ new rates $\mathbf{%
\dot{x}}^{\ast }$ and $\mathbf{\dot{\nu}}^{\ast }$\ are measured: they are
given by 
\begin{equation}
\mathbf{\dot{x}}^{\ast }=\mathbf{\dot{x}+c}\left( t\right) +\mathbf{\dot{q}}%
\left( t\right) \times \left( \mathbf{x}-\mathbf{x}_{0}\right) ,
\end{equation}%
\begin{equation}
\mathbf{\dot{\nu}}^{\ast }=\mathbf{\dot{\nu}+}\mathcal{A}\mathbf{\dot{q},}
\end{equation}%
where $\mathbf{c}$ and $\mathbf{\dot{q}}$\ are translational and rotational
velocities, respectively; $\mathcal{A}\mathbf{\dot{q}}$ is the infinitesimal
generator of the action of $SO\left( 3\right) $\ on $\mathcal{M}$ and $%
\mathcal{A}=\frac{d\mathbf{\nu }_{\mathbf{q}}}{d\mathbf{q}}\left\vert _{%
\mathbf{q}=0}\right. $ (a linear operator mapping vectors of $\mathbb{R}^{3}$%
\ into elements of the tangent space of $\mathcal{M}$) is represented by a
matrix with three columns and a number of lines equal to $\dim \mathcal{M}$,
being $\mathbf{\nu }_{\mathbf{q}}$ the order parameter measured after the
action of $SO\left( 3\right) $.

\subsection{Cracks}

We imagine that the reference place $\mathcal{B}$ be free of cracks. When a
crack is generated in the current configuration of the body, the mapping $%
\mathbf{X}\longmapsto \mathbf{x}\left( \mathbf{X}\right) $ is pointwise
one-to-one except a surface $\mathcal{C\equiv }\left\{ \mathbf{X}\in 
\mathcal{B},\text{ }f\left( \mathbf{X}\right) =0\right\} $, with $f$ a
smooth function (Fig. 1). The assumption of smoothness for $f$\ is only of
convenience. Notice that $\mathcal{C}$ is only a geometrical (non-material)
picture in the reference configuration of the real crack occurring in $%
\mathbf{x}\left( \mathcal{B}\right) $.

\FRAME{ftbpFU}{3.5276in}{2.6498in}{0pt}{\Qcb{Open crack in the current
configuration mapped by inverse motion in the reference configuration.}}{}{%
fessura1.wmf}{\special{language "Scientific Word";type
"GRAPHIC";maintain-aspect-ratio TRUE;display "USEDEF";valid_file "F";width
3.5276in;height 2.6498in;depth 0pt;original-width 9.9912in;original-height
7.491in;cropleft "0.0787579";croptop "0.9509734";cropright
"0.9120869";cropbottom "0.1175524";filename 'Fessura1.wmf';file-properties
"XNPEU";}}

The intersection of $\mathcal{C}$ with the boundary of $\mathcal{B}$ is a
regular curve $\partial \mathcal{B\cap C}$ endowed with unit normal $%
\mathfrak{m}$ such that $\mathfrak{m}\left( \mathbf{X}\right) $ belongs to
the tangent plane of $\mathcal{C}$\ at each $\mathbf{X\in \partial }\mathcal{%
B}$. Subsets $\mathfrak{b}$\ of $\mathcal{B}$ are called `\emph{parts}' here
when they are regular regions. When we consider any part $\mathfrak{b}_{%
\mathcal{C}}$ intersecting $\mathcal{C}$, the intersection $\partial 
\mathfrak{b}_{\mathcal{C}}\mathcal{\cap C}$ is a regular curve whose normal
in the tangent space of $\mathcal{C}$ is also indicated with $\mathfrak{m}$.
The normal $\mathbf{m}$ to $\mathcal{C}$ is defined by $\mathbf{m=}\frac{%
\nabla f}{\left\vert \nabla f\right\vert }$; the opposite of its surface
gradient, namely $\mathsf{L}=-\nabla _{\mathcal{C}}\mathbf{m}$, is the
curvature tensor, its trace is the opposite of the overall curvature $%
\mathcal{K}$. In the case treated here, $\mathcal{C}$ \emph{does not cross
completely }$\mathcal{B}$. The \emph{image} in $\mathcal{B}$\ of the real 
\emph{tip} of the crack is thus the margin $\mathcal{J}$ of $\mathcal{C}$
within the interior of $\mathcal{B}$. We assume that $\mathcal{J}$ is a
simple regular curve parametrized by arc length $\mathfrak{s\in }\left[ 0,%
\mathfrak{\bar{s}}\right] $ and represented by a point-valued mapping $%
\mathbf{Z:}\left[ 0,\mathfrak{\bar{s}}\right] \rightarrow \mathcal{B}$ so
that the derivative $\mathbf{Z}_{\text{,}\mathfrak{s}}\left( \mathfrak{s}%
\right) $ of $\mathbf{Z}$ with respect to $\mathfrak{s}$ is the \emph{tangent%
} vector $\mathsf{t}\left( \mathfrak{s}\right) $ at $\mathbf{Z}\left( 
\mathfrak{s}\right) $, while $\mathfrak{h}=-\mathbf{Z}_{\text{,}\mathfrak{ss}%
}$ is the curvature vector. A normal vector field $\mathsf{n}$ is chosen
along $\mathcal{J}$ to be at each $\mathbf{Z}\left( \mathfrak{s}\right) $ an
element of the tangent plane of $\mathcal{C}$ at $\mathbf{Z}\left( \mathfrak{%
s}\right) $ outward $\mathcal{C}$ (Fig. 1).

When the crack evolves in the current configuration, its picture in the
reference configuration is a\emph{\ surface} $\mathcal{C}\left( t\right) $ 
\emph{growing} in a certain time interval $\left[ 0,\bar{t}\right] $; $%
\mathcal{J}$\ has then an \emph{intrinsic fictitious relative motion} with
respect to the rest of the body, while any piece of $\mathcal{C}\left(
t\right) $\ far from $\mathcal{J}$\ remains at rest.

The intuitive behavior of the crack during the motion is simply described in
the reference place by the monotonicity of $\mathcal{C}\left( t\right) $,
namely $\mathcal{C}\left( t_{1}\right) \subseteq \mathcal{C}\left(
t_{2}\right) $, $\forall t_{1}\leq t_{2}$. We assume also that during the
time interval in which we study the motion of the crack, it does not cut
completely the body. In $\mathcal{B}$ the \emph{velocity of }$\mathcal{J}$\
is 
\begin{equation}
\mathbf{v}_{tip}=\frac{\partial \mathbf{Z}\left( \mathfrak{s},t\right) }{%
\partial t}.
\end{equation}%
Only its normal component $V=\mathbf{v}_{tip}\cdot \mathsf{n}$\ is
independent of the parametrization $\mathfrak{s}$, and in what follows, we
shall consider only $\mathbf{v}_{tip}=V\mathsf{n}$.

For any field $e\left( \mathbf{X,}t\right) $\ continuous in time and space
except $\mathcal{C}$, where it suffers bounded discontinuities, its \emph{%
jump} $\left[ e\right] $\ there is defined by the difference between the
outer and the inner trace, i.e., $\left[ e\right] =e^{+}-e^{-}$ (when the
difference makes sense), while the \emph{mean value} $\left\langle
e\right\rangle $ is given by $\left\langle e\right\rangle =\frac{1}{2}\left(
e^{+}+e^{-}\right) $. When the crack is \emph{closed} in the current
configuration, the requirement that its sides do not penetrate one into
another during the deformation is then $\left[ \mathbf{x}\right] \cdot 
\mathbf{m=}0$. Some special choices of the order parameter require the
continuity of it across $\mathcal{C}$. Without loss of generality we can
consider $\mathbf{\nu }$\ continuous across $\mathcal{C}$, while its rate
may suffer bounded discontinuities. When it is not so, since $\mathcal{M}$
is a non-linear manifold, the jump $\mathbf{\nu }^{+}-\mathbf{\nu }^{-}$
could not make sense and to define $\left[ \mathbf{\nu }\right] $ it should
be necessary to embed $\mathcal{M}$ into an appropriate linear space (a
procedure based on Withney's or Nash's theorems of embedding). In that case,
since the embedding is not unique, one should select the one convenient to
maintain the gauge properties of the underlying physics.

When there exists any field $\hat{e}\left( \mathbf{Z,}t\right) $ defined
along $\mathcal{J}\left( t\right) $ and such that $e\left( \mathbf{X,}%
t\right) \rightarrow \hat{e}\left( \mathbf{Z}\left( \mathfrak{s},t\right) 
\mathbf{,}t\right) $ as $\mathbf{X\rightarrow Z}\left( \mathfrak{s},t\right) 
$ uniformly in time, we say that $e$ has \emph{uniform limit} \emph{at the
tip} and confuse $\hat{e}\left( \mathbf{Z}\left( \mathfrak{s},t\right) 
\mathbf{,}t\right) $ with $e\left( \mathbf{Z}\left( \mathfrak{s},t\right) 
\mathbf{,}t\right) $. In this sense, we indicate the \emph{tip rate of
change of the order parameter} as $\mathbf{w}_{tip}\ $by considering it at
each $\mathbf{Z\in }\mathcal{J}$ as the uniform limit $\lim_{\mathbf{%
X\rightarrow Z}}\mathbf{\dot{\nu}}\left( \mathbf{X,}t\right) $.

\emph{Rates following the crack tip} may be defined (by chain rule) as 
\begin{equation}
\mathbf{x}^{\diamondsuit }=\mathbf{\dot{x}+Fv}_{tip},\text{ \ \ }\mathbf{\nu 
}^{\diamondsuit }=\mathbf{\dot{\nu}+}\left( \nabla \mathbf{\nu }\right) 
\mathbf{v}_{tip},
\end{equation}%
and of course the derivatives $\mathbf{x}^{\diamondsuit }$\ and\ $\mathbf{%
\nu }^{\diamondsuit }$\ are meant for points away from the tip, being rates
perceived by observers sitting on the crack tip. Their uniform limit at the
crack tip are indicated with $\mathbf{\tilde{v}}_{tip}$\ and $\mathbf{\tilde{%
w}}_{tip}$\ respectively, being $\mathbf{\tilde{v}}_{tip}$ the velocity of
the deformed tip and $\mathbf{\tilde{w}}_{tip}$ the rate of $\mathbf{\nu }$\
at the deformed tip.

Moreover, let $\mathfrak{b}$\ be any part of $\mathcal{B}$. The boundary $%
\partial \mathfrak{b}$\ of $\mathfrak{b}$\ is a two-dimensional surface (of
normal $\mathbf{n}$) and may be parametrized by parameters $u_{1}$ and $u_{2}
$. If we consider $\mathfrak{b}$ varying in time, i.e., $\mathfrak{b}\left(
t\right) $, points $\mathbf{X}$ of the boundary $\partial \mathfrak{b}\left(
t\right) $ are identified by $\mathbf{X}\left( u_{1},u_{2},t\right) $ so
that the velocity $\mathbf{u}$ of $\partial \mathfrak{b}\left( t\right) $\
is given by 
\begin{equation}
\mathbf{u=}\frac{\partial \mathbf{X}\left( u_{1},u_{2},t\right) }{\partial t}%
.
\end{equation}%
Only the normal component $U=\mathbf{u\cdot n}$ is independent of the
parametrization $\left( u_{1},u_{2}\right) $. \emph{Rates following the
moving boundary} $\partial \mathfrak{b}\left( t\right) $\ are then given by%
\begin{equation}
\mathbf{x}^{\circ }=\mathbf{\dot{x}+Fu,}\text{ \ \ }\mathbf{\nu }^{\circ }=%
\mathbf{\dot{\nu}+}\left( \nabla \mathbf{\varphi }\right) \mathbf{u.}
\end{equation}

\section{Balance of standard and substructural interactions via an
invariance argument}

In the standard mechanics of deformable bodies, common stresses and bulk
forces are the sole measures of interaction. When the material substructure
is accounted for, the picture of the interactions become more articulated
because we must consider objects measuring the extra power developed in the
rate $\mathbf{\dot{\nu}}$.

Below, we summarize balance equations for standard and substructural
interactions in the bulk, at the lateral margins of the crack and at the tip
(and the attention is focused on the substructural ones because the others
are well known). To obtain them we could follow two ways: (i) we could use
variational arguments involving a Lagrangian and its invariance with respect
to general Lie groups (underlining in this way their covariance\footnote{%
Following in this way the guidelines of the basic proof of covariance
developed for simple materials in (Marsden \& Hughes, 1983).}) or (ii) we
could require $SO\left( 3\right) $ invariance of the power of `external'
interactions. Though we prefer the former, we sketch here the latter
because, in this way, we may clearly underline the distinction between
balance equations and the constitutive structure of the interactions
involved (two aspects that are mixed in an Hamiltonian approach). Such a
distinction clarifies our aim to obtain relations valid even for general
irreversible processes in deformable solids in different circumstances,
although our treatment deals mainly with non-linear elastic processes,
unless otherwise stated.

\subsection{Balance of interactions in the bulk}

Let $\mathfrak{b}$\ be any arbitrary part of $\mathcal{B}$ (of boundary $%
\partial \mathfrak{b}$)\ far from the crack. The external power $\mathcal{P}%
_{\mathfrak{b}}^{ext}$ of the standard and substructural interactions on $%
\mathfrak{b}$ is a linear real functional on the pairs $\left( \mathbf{\dot{x%
},\dot{\nu}}\right) $ represented by 
\begin{equation}
\mathcal{P}_{\mathfrak{b}}^{ext}\left( \mathbf{\dot{x},\dot{\nu}}\right)
=\int_{\mathfrak{b}}\left( \mathbf{b\cdot \dot{x}+\beta \cdot \dot{\nu}}%
\right) +\int_{\partial \mathfrak{b}}\left( \mathbf{Tn\cdot \dot{x}+}%
\mathcal{S}\mathbf{n\cdot \dot{\nu}}\right) ,
\end{equation}%
where the substructural interactions are measured through volume $\mathbf{%
\beta }$ and surface $\mathcal{S}\mathbf{n}$ densities as in the case of
standard interactions. The bulk density $\mathbf{\beta }$\ may account for
both possible substructural inertia effects and interactions due, e.g., to
electromagnetic fields acting on the substructure (as in the case of
ferroelectrics). Both $\mathbf{b}$ and $\mathbf{\beta }$ are continuous on $%
\mathfrak{b}$. $\mathcal{S}$ is called \emph{microstress} and maps linearly
vectors of $\mathbb{R}^{3}$ into elements of the cotangent space of $%
\mathcal{M}$. Roughly speaking, $\mathbf{\tau }=\mathcal{S}\mathbf{n}$ is a
`generalized traction'; the product $\mathbf{\tau \cdot \dot{\nu}}$ is the
power exchanged between two adjacent parts at $\mathbf{X}$ through a surface
of normal $\mathbf{n}$, as a consequence of the change of the substructure
at the same point. Since our analysis is developed in the reference place $%
\mathcal{B}$, the first Piola-Kirchhoff stress $\mathbf{T}$ is used: it
associates tensions in $\mathbf{x}\left( \mathcal{B}\right) $ to vectors in $%
\mathcal{B}$\ and is the pull-back in $\mathcal{B}$ of the standard (Cauchy)
stress $\mathbf{\sigma }$ measuring "true" tensions in the current place $%
\mathbf{x}\left( \mathcal{B}\right) $ of the body: in fact, $\mathbf{T=}%
\left( \det \mathbf{F}\right) \mathbf{\sigma F}^{-T}$. Accordingly, the
co-vector $\mathbf{b}$ is the pull back of the body forces (including
inertia) living in the current place $\mathbf{x}\left( \mathcal{B}\right) $.

We impose that $\mathcal{P}_{\mathfrak{b}}^{ext}$ \emph{is invariant under SO%
}$\left( 3\right) $, \emph{for any }$\mathfrak{b}$, i.e. 
\begin{equation}
\mathcal{P}_{\mathfrak{b}}^{ext}\left( \mathbf{\dot{x}}^{\ast }\mathbf{,\dot{%
\nu}}^{\ast }\right) =\mathcal{P}_{\mathfrak{b}}^{ext}\left( \mathbf{\dot{x},%
\dot{\nu}}\right) ,
\end{equation}%
\emph{for any choice of} $\mathbf{c}\left( t\right) $, $\mathbf{\dot{q}}%
\left( t\right) $ \emph{and} $\mathfrak{b}$.

By using (1) and (2), thanks to the arbitrariness of $\mathbf{c}$\ and $%
\mathbf{\dot{q}}$, we obtain the \emph{standard integral balance of forces} 
\begin{equation}
\int_{\mathfrak{b}}\mathbf{b}+\int_{\partial \mathfrak{b}}\mathbf{Tn=0,}
\end{equation}%
and a \emph{generalized integral balance of moments}:%
\begin{equation}
\int_{\mathfrak{b}}\left( \left( \mathbf{x-x}_{0}\right) \times \mathbf{b+}%
\mathcal{A}^{T}\mathbf{\beta }\right) +\int_{\partial \mathfrak{b}}\left(
\left( \mathbf{x-x}_{0}\right) \times \mathbf{Tn+}\mathcal{A}^{T}\mathcal{S}%
\mathbf{n}\right) =0.
\end{equation}%
From (9) the common pointwise balance 
\begin{equation}
\mathbf{b}+Div\mathbf{T=0}\text{ \ \ \ \ \ \ \ in }\mathcal{B}
\end{equation}%
follows thanks to the arbitrariness of $\mathfrak{b}$, while, from (11), we
get 
\begin{equation}
\mathcal{A}^{T}\left( Div\mathcal{S}+\mathbf{\beta }\right) =\mathsf{e}%
\mathbf{TF}^{T}\mathbf{-}\left( \nabla \mathcal{A}^{T}\right) \mathcal{S}%
\mathbf{.}
\end{equation}%
with $\mathsf{e}$ Ricci's alternator. The condition (12) implies that the
co-vector $\mathsf{e}_{ABC}T_{i}^{B}F_{i}^{C}-\left( \nabla _{A}\mathcal{A}%
_{B}^{\alpha }\right) \mathcal{S}_{\alpha }^{B}$ \footnote{$A,B...$ denote
components in $\mathcal{B}$, $i,j...$ components in $\mathbf{x}\left( 
\mathcal{B}\right) $ and $\alpha ...$ components over the atlas on $\mathcal{%
M}$.} belongs to the range of the linear operator $\mathcal{A}^{T}$ at each $%
\mathbf{\nu }$. However, $\mathcal{A}$ is \emph{not one-to-one}, then at
each $\mathbf{\nu }$\ there may exist an element $\mathbf{z}$ of the
cotangent space of $\mathcal{M}$ at $\mathbf{\nu }$ such that 
\begin{equation}
\mathcal{A}^{T}\mathbf{z=}\mathsf{e}\mathbf{TF}^{T}-\left( \nabla \mathcal{A}%
^{T}\right) \mathcal{S},
\end{equation}%
which implies 
\begin{equation}
Div\mathcal{S}-\mathbf{z+\beta =0}\text{\ \ \ \ \ \ \ in }\mathcal{B}.
\end{equation}

Equation (14) is the \emph{pointwise balance of substructural interactions}
(Capriz 1989) and $\mathbf{z}$\ is an internal \emph{self-force}. Equation
(13) states that the presence of substructural interactions renders
unsymmetrical the Cauchy stress $\mathbf{\sigma }$\ given by $\mathbf{\sigma
=}\left( \det \mathbf{F}\right) ^{-1}\mathbf{TF}^{T}$. The case of scalar
order parameters seems to be pathological for the procedure used here
because $\mathcal{A}$ vanishes. Formally, one may circumvent the problem by
making use of spherical second order tensors, obtaining the balance (14),
then reducing it to the scalar case (which would be in any case the
Euler-Lagrange equation of some Lagrangian). If one is skeptical about such
an interpretation, one could accept the point of view discussed here to
derive (14) only for other types of order parameters and postulate (14) a
priori in the scalar case.

\subsection{Balance of interactions along the sides of the crack}

Let $\mathfrak{b}_{\mathcal{C}}$ be any arbitrary part intersecting the
crack $\mathcal{C}$ far from the tip. If one writes $\mathcal{P}_{\mathfrak{b%
}_{\mathcal{C}}}^{ext}$ and requires $SO\left( 3\right) $ invariance, the
integral balances (9) and (10) follow but $\mathfrak{b}_{\mathcal{C}}$\ is
now the domain of integration. If we shrink $\mathfrak{b}_{\mathcal{C}}$\ to 
$\mathfrak{b}_{\mathcal{C}}\cap \mathcal{C}$, since the integrands of the
volume integrals are continuous while the stress does not, we obtain%
\begin{equation}
\int_{\mathfrak{b}_{\mathcal{C}}}\mathbf{b\rightarrow 0}\text{ \ \ \ \ as }%
\mathfrak{b}_{\mathcal{C}}\rightarrow \mathfrak{b}_{\mathcal{C}}\cap 
\mathcal{C},
\end{equation}%
\begin{equation}
\int_{\partial \mathfrak{b}_{\mathcal{C}}}\mathbf{Tn\rightarrow }\int_{%
\mathfrak{b}_{\mathcal{C}}\cap \mathcal{C}}\left[ \mathbf{T}\right] \mathbf{m%
}\text{ \ \ \ \ as }\mathfrak{b}_{\mathcal{C}}\rightarrow \mathfrak{b}_{%
\mathcal{C}}\cap \mathcal{C}.
\end{equation}%
Consequently, the arbitrariness of $\mathfrak{b}_{\mathcal{C}}$ implies from
(9) the common pointwise balance 
\begin{equation}
\left[ \mathbf{T}\right] \mathbf{m=0}\text{ \ \ \ \ along }\mathcal{C},
\end{equation}%
An analogous reasoning can be applied to (10) and leads to 
\begin{equation}
\int_{\partial \mathfrak{b}_{\mathcal{C}}}\mathcal{A}^{T}\left[ \mathcal{S}%
\right] \mathbf{n=0}
\end{equation}%
as $\mathfrak{b}_{\mathcal{C}}\rightarrow \mathfrak{b}_{\mathcal{C}}\cap 
\mathcal{C}$. The arbitrariness of $\mathfrak{b}_{\mathcal{C}}$ implies the
pointwise balance $\mathcal{A}^{T}\left[ \mathcal{S}\right] \mathbf{m=0}$\
which is tantamount to write 
\begin{equation}
\left[ \mathcal{S}\right] \mathbf{m=z}_{\mathcal{C}}^{\prime }\text{ \ \ \ \
along }\mathcal{C}\text{, \ \ with }\mathcal{A}^{T}\mathbf{z}_{\mathcal{C}%
}^{\prime }=0.
\end{equation}%
(\emph{balance of substructural interactions along the sides of the crack}). 
$\mathbf{z}_{\mathcal{C}}^{\prime }$ vanishes when at each $\mathbf{\nu }$
the range of $\mathcal{A}$ covers the whole tangent space of $\mathcal{M}$
there.

\subsection{Balance of interactions at the tip of the crack}

Take a part of $\mathcal{B}$ with the form of a `curved cylinder' (see,
e.g., Fig. 1) $\mathfrak{b}_{R}^{\ast }$ obtained by translating a disc $%
D_{R}$ of diameter $R$ from $\mathbf{Z}\left( \mathfrak{s}_{1}\right) $ to $%
\mathbf{Z}\left( \mathfrak{s}_{2}\right) $\ (two arbitrary points of $%
\mathcal{J}$\ with $\mathfrak{s}_{1}<\mathfrak{s}_{2}$) maintaining $D_{R}$\
orthogonal to $\mathsf{t}\left( \mathfrak{s}\right) $\ at each $\mathbf{Z}%
\left( \mathfrak{s}\right) $\ and the centre of $D_{R}$\ coincident with $%
\mathbf{Z}\left( \mathfrak{s}\right) $\ at\ each $\mathfrak{s}$. The
external power of all interactions acting over $\mathfrak{b}_{R}^{\ast }$\ is%
\begin{equation*}
\mathcal{P}_{\mathfrak{b}_{R}^{\ast }}^{ext}\left( \mathbf{\dot{x},\dot{\nu}}%
,\mathbf{\tilde{v}}_{tip},\mathbf{\tilde{w}}_{tip}\right) =\int_{\mathfrak{b}%
_{R}^{\ast }}\left( \mathbf{b\cdot \dot{x}+\beta \cdot \dot{\nu}}\right) +
\end{equation*}%
\begin{equation}
+\int_{\partial \mathfrak{b}_{R}^{\ast }}\left( \mathbf{Tn\cdot \dot{x}+}%
\mathcal{S}\mathbf{n\cdot \dot{\nu}}\right) +\int_{\mathfrak{s}_{1}}^{%
\mathfrak{s}_{2}}\left( \mathbf{b}_{tip}\cdot \mathbf{\tilde{v}}_{tip}+%
\mathbf{\beta }_{tip}\mathbf{\cdot \tilde{w}}_{tip}\right) ,
\end{equation}%
where $\mathbf{b}_{tip}$\ collects \emph{only the inertia effects at the tip}%
, and $\mathbf{\beta }_{tip}$\ accounts formally for \emph{possible
substructural inertia effects at the tip}.

The changes of observers described in (1) and (2) can be written for the
velocities at the tip as 
\begin{equation}
\mathbf{\tilde{v}}_{tip}^{\ast }=\mathbf{\tilde{v}}_{tip}+\mathbf{c}\left(
t\right) +\mathbf{\dot{q}}\left( t\right) \times \left( \mathbf{x}_{tip}-%
\mathbf{x}_{0}\right) ,
\end{equation}%
\begin{equation}
\mathbf{\tilde{w}}_{tip}^{\ast }=\mathbf{\tilde{w}}_{tip}\mathbf{+}\mathcal{A%
}_{tip}\mathbf{\dot{q},}
\end{equation}%
where $\mathcal{A}_{tip}=\frac{d\mathbf{\nu }_{\mathbf{q}}\left( \mathbf{Z}%
\right) }{d\mathbf{q}}\left\vert _{\mathbf{q}=0}\right. $ in the sense of
uniform limit. We impose here $SO\left( 3\right) $\ invariance requiring
that 
\begin{equation}
\mathcal{P}_{\mathfrak{b}_{R}^{\ast }}^{ext}\left( \mathbf{\dot{x}}^{\ast }%
\mathbf{,\dot{\nu}}^{\ast },\mathbf{\tilde{v}}_{tip}^{\ast },\mathbf{\tilde{w%
}}_{tip}^{\ast }\right) =\mathcal{P}_{\mathfrak{b}_{R}^{\ast }}^{ext}\left( 
\mathbf{\dot{x},\dot{\nu},\tilde{v}}_{tip},\mathbf{\tilde{w}}_{tip}\right) ,
\end{equation}%
for any choice of $\mathbf{c}\left( t\right) $, $\mathbf{\dot{q}}\left(
t\right) $ and $\mathfrak{b}_{R}^{\ast }$. From (23), thanks to the
arbitrariness of $\left[ \mathfrak{s}_{1},\mathfrak{s}_{2}\right] $, by
shrinking $\mathfrak{b}_{R}^{\ast }$\ to the tip $\left( R\rightarrow
0\right) $, we obtain%
\begin{equation}
\mathbf{b}_{tip}+\int_{tip}\mathbf{Tn}=0
\end{equation}%
(\emph{balance of standard forces at the tip}),%
\begin{equation}
\mathbf{\beta }_{tip}+\int_{tip}\mathcal{S}\mathbf{n=z}_{tip}^{\prime },%
\text{ \ \ with }\mathcal{A}_{tip}^{T}\mathbf{z}_{tip}^{\prime }=0
\end{equation}%
(\emph{balance of substructural interactions at the tip}), where we use the
notation $\int_{tip}\left( \cdot \right) $ for $\lim_{R\rightarrow
0}\int_{D_{R}}\left( \cdot \right) $, interpreting it in the sense of
uniform limits.

\subsection{Standard and substructural inertia effects}

\subsubsection{Bulk terms}

The bulk interactions $\mathbf{b}$ and $\mathbf{\beta }$\ contain both
inertial and non-inertial terms, namely $\mathbf{b=b}^{in}\mathbf{+b}^{ni}$
and $\mathbf{\beta =\beta }^{in}\mathbf{+\beta }^{ni}$. The overall power of
the inertial terms over any stationary part $\mathfrak{b}$\ far from the
crack is equal to the opposite of the rate of kinetic energy of $\mathfrak{b}
$, the density of which is commonly chosen to be the sum $\rho \left( \frac{1%
}{2}\left\Vert \mathbf{\dot{x}}\right\Vert ^{2}+k\left( \mathbf{\nu },%
\mathbf{\dot{\nu}}\right) \right) $, where $k\left( \mathbf{\nu },\mathbf{%
\dot{\nu}}\right) $\ the kinetic contribution of the substructure, if there
is some experimental evidence of it. From the arbitrariness of $\mathfrak{b}$%
, the common identification of the inertial terms follows (Capriz, 1989): 
\begin{equation}
\mathbf{b}^{in}=-\rho \mathbf{\ddot{x},}\text{ \ \ }\mathbf{\beta }^{in}=-%
\frac{d}{dt}\rho \partial _{\mathbf{\dot{\nu}}}\chi +\rho \partial _{\mathbf{%
\nu }}\chi ,
\end{equation}%
where $\chi $\ is the \emph{substructural kinetic co-energy density}: its
Legendre transform with respect to the rate coincides with the kinetic
energy $k\left( \mathbf{\nu },\mathbf{\dot{\nu}}\right) $. In common cases,
the experiments show that the term $\frac{d}{dt}\rho \partial _{\mathbf{\dot{%
\nu}}}\chi -\rho \partial _{\mathbf{\nu }}\chi $\ is in general negligible,
unless the substructure oscillates at very high frequencies.

\subsubsection{Tip effects}

The standard results before allow us to identify explicitly the tip inertial
terms. We consider a part around the tip the `curved cylinder' $\mathfrak{b}%
_{R}^{\ast }$ varying it in time, $\mathfrak{b}_{R}^{\ast }\left( t\right) $%
, to follow the growth of the crack. We then write the integral versions of
(26) over $\mathfrak{b}_{R}^{\ast }$\ adding not only a \emph{tip inertial
term} $\mathbf{b}_{tip}$, as in the common treatments, but also a \emph{%
substructural tip inertial term } $\mathbf{\beta }_{tip}$, and obtaining then%
\begin{equation}
\int_{\mathfrak{b}_{R}^{\ast }\left( t\right) }\mathbf{b}^{in}+\int_{%
\mathfrak{s}_{1}}^{\mathfrak{s}_{2}}\mathbf{b}_{tip}=-\frac{d}{dt}\int_{%
\mathfrak{b}_{R}^{\ast }\left( t\right) }\rho \mathbf{\dot{x}+}%
\int_{\partial \mathfrak{b}_{R}^{\ast }\left( t\right) }\rho \mathbf{\dot{x}}%
U,
\end{equation}%
\begin{equation*}
\int_{\mathfrak{b}_{R}^{\ast }\left( t\right) }\mathbf{\beta }^{in}+\int_{%
\mathfrak{s}_{1}}^{\mathfrak{s}_{2}}\mathbf{\beta }_{tip}=-\frac{d}{dt}\int_{%
\mathfrak{b}_{R}^{\ast }\left( t\right) }\rho \partial _{\mathbf{\dot{\nu}}%
}\chi +
\end{equation*}%
\begin{equation}
+\int_{\mathfrak{b}_{R}^{\ast }\left( t\right) }\rho \partial _{\mathbf{\nu }%
}\chi +\int_{\partial \mathfrak{b}_{R}^{\ast }\left( t\right) }\rho \partial
_{\mathbf{\dot{\nu}}}\chi U.
\end{equation}%
The last integral in (27) is the inflow of standard momentum through the
boundary $\partial \mathfrak{b}_{R}^{\ast }\left( t\right) $\ of $\mathfrak{b%
}_{R}^{\ast }$, due to the `virtual' (non-material) movement of $\mathfrak{b}%
_{R}^{\ast }$\ in time; an analogous meaning has the last integral in (28).
By shrinking $\mathfrak{b}_{R}^{\ast }$ at the tip $\mathfrak{I}$ (letting $%
R\rightarrow 0$), we get 
\begin{equation}
\mathbf{b}_{tip}=\int_{tip}\rho \mathbf{\dot{x}}\left( \mathbf{v}_{tip}\cdot 
\mathbf{n}\right) ,
\end{equation}%
\begin{equation}
\mathbf{\beta }_{tip}=\int_{tip}\rho \partial _{\mathbf{\dot{\nu}}}\chi
\left( \mathbf{\nu },\mathbf{\dot{\nu}}\right) \left( \mathbf{v}_{tip}\cdot 
\mathbf{n}\right) ,
\end{equation}%
thanks to the arbitrariness of $\left[ \mathfrak{s}_{1},\mathfrak{s}_{2}%
\right] $. The obvious changes in (24) and (25) follow. The result (29) is
standard in the theory of dynamic fracture in simple bodies.

\subsection{Stresses at the tip}

The speed of the crack growth is finite, so we may assume that $%
\int_{\partial D_{R}}\mathbf{n}\otimes \rho \mathbf{\dot{x}}$ be bounded up
to the tip as $R\rightarrow 0$. Such an assumption (which is in certain
sense on the behavior of the solution) implies that

\begin{equation}
\int_{tip}\mathbf{n}\otimes \rho \mathbf{\dot{x}=0.}
\end{equation}%
Moreover, we assume also that the tip flow of substructural momentum be
bounded up to the tip, i.e., $\int_{\partial D_{R}}\rho \partial _{\mathbf{%
\dot{\nu}}}\chi U$ \ is bounded as $R\rightarrow 0$ for any choice of the
order parameter. This implies that $\partial _{\mathbf{\dot{\nu}}}\chi $\ is
bounded as $R\rightarrow 0$\ (because $U$ is bounded) and then 
\begin{equation}
\int_{tip}\mathbf{n}\otimes \rho \partial _{\mathbf{\dot{\nu}}}\chi =\mathbf{%
0.}
\end{equation}

If the range of $\mathcal{A}_{tip}$\ covers the whole tangent space of $%
\mathcal{M}$ at $\mathbf{\nu }\left( \mathbf{Z},t\right) $, we get $\mathbf{z%
}_{tip}^{\prime }=\mathbf{0}$. Consequently, from (24), (25), (29), (30), we
get

\begin{equation}
\text{ \ \ }\int_{tip}\mathbf{Tn=0,}\text{ \ \ }\int_{tip}\mathcal{S}\mathbf{%
n=0.}
\end{equation}

We could reverse the point of view, following a remark of Landau and Lif\v{s}%
its, and we could say that since we allow the crack to evolve, the stresses
are bounded up to the tip. This circumstance together with the assumption
(32) would imply (33) directly and (31) and $\mathbf{z}_{tip}^{\prime }=%
\mathbf{0}$ as further consequences.

\section{Interactions due to the growth of the crack}

The evolution of the crack is represented by the `fictitious' growth of $%
\mathcal{C}$ in the reference configuration $\mathcal{B}$ (which on the
contrary would remain fixed once and for all). Such a growth generates an
independent kinematics in $\mathcal{B}$ and interactions power-conjugated
with it in the bulk, at the lateral sides and at the tip of the crack. They
should satisfy appropriate balances. These interactions have a twofold
nature: from one hand they live in $\mathcal{B}$ and are thus different from
the standard and substructural interactions (notice, e.g., that (11) is the
balance of forces living in $\mathbf{x}\left( \mathcal{B}\right) $); from
the other hand, since the kinematics of $\mathcal{C}$ is only `apparent'
(non-material) in the reference configuration, the new `forces' may be
expressed in terms of standard and substructural interactions and the free
energy. We list below these `fictitious forces' and their balances that are
commonly justified in various manners and used to describe different cases
of mutations in bodies. However, we do not give contribution to the current
discussion about the attribution of such balances \footnote{%
See the results and the theoretical discussions in (Abeyaratne \& Knowles,
1990; Eshelby, 1975; Epstein, 2002; Gurtin, 1995; James, 2002; Segev, 1996; 
\v{S}ilhav\'{y}, 1997).}. In the identification of them in terms of standard
and substructural interactions, our contribution relies in the deduction of
the substructural components, and this step is crucial toward the main
result of the present paper.

\subsection{Balance in the bulk of forces due to the crack growth}

For any time varying part $\mathfrak{b}\left( t\right) $ (remind that the
evolution of $\mathfrak{b}$ is not material) we consider bulk internal and
external forces, namely the vectors $\mathbf{g}$ and $\mathbf{e}$
respectively, and a stress $\mathbb{P}$, a second order tensor that maps at
each $\mathbf{X}$ the tangent space there onto the corresponding cotangent
space. It is commonly postulated that they satisfy the integral balance%
\begin{equation}
\int_{\partial \mathfrak{b}\left( t\right) }\mathbb{P}\mathbf{n+}\int_{%
\mathfrak{b}\left( t\right) }\left( \mathbf{g+e}\right) =0,
\end{equation}%
for any choice of $\mathfrak{b}\left( t\right) $. The pointwise balance%
\begin{equation}
Div\mathbb{P+}\mathbf{g+e=0,}\text{ \ \ \ \ \ \ in }\mathcal{B},
\end{equation}%
then follows.

In subsequent steps, the strategy foresees the identification of the various
elements of (35) in terms of the true stresses by making use of a mechanical
dissipation inequality (a mechanical version of the second law). We use such
a procedure having in mind the need to manage an approach valid even in
fully dissipative processes like viscosity or plastic flows. However, we
develop the identification of the terms of (35) just in non-linear
elasticity (suggesting also the necessary developments in other
circumstances) to put in evidence the basic fact that in that case, the
balance (35) reduces to one of the conservation laws that can be deduced
from a N\"{o}ther-like theorem, precisely the one associated with the
invariance of the Lagrangian with respect to diffeomorphisms altering $%
\mathcal{B}$.

\subsection{Balance along the sides of the crack of forces due to the crack
growth}

For any arbitrary part $\mathfrak{b}_{\mathcal{C}}\left( t\right) $\
intersecting the crack far from the tip, we write the balance (34) adding 
\emph{surface measures of interaction }along the margins of the crack,
namely a surface stress $\bar{\sigma}\left( \mathbf{I-m\otimes m}\right) $,
with $\mathbf{I}$ the unit second order tensor and $\bar{\sigma}\left( 
\mathbf{X}\right) $ a scalar function continuous up to the tip where it is
indicated with $\bar{\sigma}_{tip}$; and an internal surface force $\mathbf{g%
}_{\mathcal{C}}$ (vector). So that we obtain%
\begin{equation}
\int_{\partial \mathfrak{b}_{\mathcal{C}}\left( t\right) }\mathbb{P}\mathbf{%
n+}\int_{\mathfrak{b}_{\mathcal{C}}\left( t\right) }\left( \mathbf{g+e}%
\right) +\int_{\left( \partial \mathfrak{b}_{\mathcal{C}}\left( t\right)
\right) \cap \mathcal{C}\left( t\right) }\bar{\sigma}\mathfrak{m}\mathbf{+}%
\int_{\mathfrak{b}_{\mathcal{C}}\left( t\right) \cap \mathcal{C}\left(
t\right) }\mathbf{g}_{\mathcal{C}}.
\end{equation}%
The arbitrariness of $\mathfrak{b}_{\mathcal{C}}$\ and the bulk balance (35)
imply%
\begin{equation}
\left[ \mathbb{P}\right] \mathbf{m+g}_{\mathcal{C}}+\nabla _{\mathcal{C}}%
\bar{\sigma}+\bar{\sigma}\left( \mathcal{K}\mathbf{I}-\text{\textsf{L}}%
\right) \mathbf{m=0}\text{ \ \ \ \ \ along }\mathcal{C},
\end{equation}%
where $\nabla _{\mathcal{C}}$ denotes the surface gradient (see Section 2).\ 

\subsection{Balance of tip forces due to the crack growth}

Along the tip we consider a line tension $\lambda _{tip}\mathsf{t}$, a tip
internal force $\mathbf{g}_{tip}$ (vector), and a tip external \emph{inertial%
}\ force $\mathbf{e}_{tip}$ (vector). If we consider a `curved cylinder' $%
\mathfrak{b}_{R}^{\ast }\left( t\right) $ (of the type used above) around
the tip, intersecting $\mathcal{J}$ in two points, say $\mathbf{Z}\left( 
\mathfrak{s}_{1}\right) $ and $\mathbf{Z}\left( \mathfrak{s}_{2}\right) $,
in writing on $\mathfrak{b}_{R}^{\ast }\left( t\right) $\ the balance of
interactions power-conjugated with the `fictitious' kinematics of $\mathcal{C%
}\left( t\right) $, we add a line term%
\begin{equation}
\left( \lambda _{tip}\mathsf{t}\left( \mathfrak{s}_{2}\right) -\lambda _{tip}%
\mathsf{t}\left( \mathfrak{s}_{1}\right) \right) \mathbf{+}\int_{\mathfrak{s}%
_{1}}^{\mathfrak{s}_{2}}\left( \mathbf{g}_{tip}\mathbf{+e}_{tip}\right) 
\mathbf{.}
\end{equation}%
to the bulk and surface terms used in (36). By shrinking $\mathfrak{b}%
_{R}^{\ast }\left( t\right) $ up to the tip letting $R\rightarrow 0$
(uniformly in time), the arbitrariness of $\left[ \mathfrak{s}_{1},\mathfrak{%
s}_{2}\right] $ implies \footnote{%
The terms $\nabla _{\mathcal{C}}\bar{\sigma}+\bar{\sigma}\left( \mathcal{K}%
\mathbf{I}-\text{\textsf{L}}\right) \mathbf{m}$ and $\mathbf{e}_{tip}$\ mark
the difference with the equations similar to (35), (37) and (39) discussed
for the quasi-static evolution of planar three-dimensional cracks in (Gurtin
\& Shvartsman, 1997).} 
\begin{equation}
\mathbf{g}_{tip}\mathbf{+e}_{tip}-\bar{\sigma}_{tip}\mathsf{n}-\lambda _{tip}%
\mathfrak{h}+\int_{tip}\mathbb{P}\mathbf{n}=0\text{ \ \ \ \ \ \ along }%
\mathfrak{I}.
\end{equation}

\subsection{Identification of the inertial term $\mathbf{e}_{tip}$}

To identify explicitly $\mathbf{e}_{tip}$ in terms of the standard and
substructural measures of interaction, it suffices to consider a `curved
cylinder' $\mathfrak{b}_{R}^{\ast }\left( t\right) $\ wrapped around the tip
and to write an inertial balance of the type 
\begin{equation*}
\mathbb{K}^{rate}\left( \mathfrak{b}_{R}^{\ast }\left( t\right) \right)
+\int_{\mathfrak{b}_{R}^{\ast }\left( t\right) }\left( \mathbf{b}^{in}%
\mathbf{\cdot \dot{x}+\beta }^{in}\mathbf{\cdot \dot{\nu}}\right) +
\end{equation*}%
\begin{equation}
+\int_{\mathfrak{s}_{1}}^{\mathfrak{s}_{2}}\left( \mathbf{b}_{tip}\cdot 
\mathbf{\tilde{v}}_{tip}+\mathbf{\beta }_{tip}\mathbf{\cdot \tilde{w}}_{tip}+%
\mathbf{e}_{tip}\cdot \mathbf{v}_{tip}\right) =0,
\end{equation}%
where $\mathbb{K}^{rate}\left( \mathfrak{b}_{R}^{\ast }\left( t\right)
\right) $\ is the rate of the kinetic energy of $\mathfrak{b}_{R}^{\ast
}\left( t\right) $\ and, of course, $\mathbf{b}_{tip}\ $and $\mathbf{\beta }%
_{tip}$ develop power in the actual rates. Since $\mathfrak{b}_{R}^{\ast }$\
varies in time, $\mathbb{K}^{rate}$\ is the difference between the time
derivative of the integral of $\rho \left( \frac{1}{2}\left\Vert \mathbf{%
\dot{x}}\right\Vert ^{2}+k\left( \mathbf{\nu },\mathbf{\dot{\nu}}\right)
\right) $ and its inflow through the moving boundary $\partial \mathfrak{b}%
_{R}^{\ast }\left( t\right) .$ By shrinking $\mathfrak{b}_{R}^{\ast }$\ up
to the tip (uniformly in time), the arbitrariness of $\left[ \mathfrak{s}%
_{1},\mathfrak{s}_{2}\right] $\ and the use of the identities (26), (29),
(30) and (32) imply%
\begin{equation}
\mathbf{e}_{tip}=\int_{tip}\rho \left( \frac{1}{2}\left\Vert \mathbf{\dot{x}}%
\right\Vert ^{2}+k\left( \mathbf{\nu },\mathbf{\dot{\nu}}\right) \right) 
\mathbf{n.}
\end{equation}

\section{The mechanical dissipation inequality and its consequences}

Consider $\nabla \mathbf{\nu }$ and $\mathcal{S}$. We define the product%
\emph{\ }$\underline{\ast }$\emph{\ }by 
\begin{equation}
\left( \nabla \mathbf{\nu }^{T}\underline{\ast }\mathcal{S}\right) \mathbf{%
n\cdot u=}\mathcal{S}\mathbf{n\cdot \left( \nabla \mathbf{\nu }\right) u,}
\end{equation}%
for any pair of vectors\emph{\ }$\mathbf{n}$ and $\mathbf{u}$. Some special
cases are the following: when $\mathbf{\nu }$\ is \emph{scalar}, $\nabla 
\mathbf{\nu }^{T}\underline{\ast }\mathcal{S}=\nabla \mathbf{\nu }\otimes 
\mathcal{S}$, when $\mathbf{\varphi }$\ is a \emph{vector}, $\nabla \mathbf{%
\nu }^{T}\underline{\ast }\mathcal{S}=\nabla \mathbf{\nu }^{T}\mathcal{S}$,
while when $\mathbf{\nu }$\ is a \emph{second order tensor}, $\nabla \mathbf{%
\nu }^{T}\underline{\ast }\mathcal{S}=\nabla \mathbf{\nu }^{T}:\mathcal{S}$.

\subsection{The formal statement of the mechanical dissipation inequality}

An isothermal version of such an inequality is the following: 
\begin{equation}
\Psi _{\mathfrak{b}}^{rate}-\mathcal{P}_{\mathfrak{b}}^{ext}\leq 0,
\end{equation}%
where $\Psi _{\mathfrak{b}}^{rate}$\ is the rate of the Helmholtz free
energy of $\mathfrak{b}$ the power of interactions over $\mathfrak{b}$. The
standard assumption is that \emph{the (here purely mechanical) inequality }%
(43)\emph{\ holds for any choice of the rates involved and for any part }$%
\mathfrak{b}$.

Below we account for time varying parts $\mathfrak{b}\left( t\right) $ of $%
\mathcal{B}$ to follow in $\mathcal{B}_{0}$ the growth of the crack in $%
\mathcal{B}$, then the external power $\mathcal{P}_{\mathfrak{b}}^{ext}$
must account for the interactions associated with such a growth.

\subsection{The mechanical dissipation inequality in the bulk. Consequences.}

Let $\mathfrak{b}\left( t\right) $\ be any time varying part of $\mathcal{B}$
far from the crack. $\Psi _{\mathfrak{b}}$\ is expressed only by means of a
bulk free energy density $\psi $\ and we have 
\begin{equation*}
\int_{\mathfrak{b}\left( t\right) }\dot{\psi}+\int_{\partial \mathfrak{b}%
\left( t\right) }\psi \left( \mathbf{u\cdot n}\right) -\int_{\mathfrak{b}%
\left( t\right) }\left( \mathbf{b\cdot \dot{x}+\beta \cdot \dot{\nu}}\right)
-
\end{equation*}%
\begin{equation}
-\int_{\partial \mathfrak{b}\left( t\right) }\left( \mathbf{Tn\cdot \dot{x}}+%
\mathcal{S}\mathbf{n\cdot \dot{\nu}}+\left( \mathbb{P+}\mathbf{F}^{T}\mathbf{%
T+}\nabla \mathbf{\nu }^{T}\underline{\ast }\mathcal{S}\right) \mathbf{n}%
\cdot \mathbf{u}\right) \leq 0.
\end{equation}%
where we have used (6) to transform the original integrand $\mathbf{Tn\cdot x%
}^{\circ }+\mathcal{S}\mathbf{n\cdot \nu }^{\circ }+\mathbb{P}\mathbf{n}%
\cdot \mathbf{u}$ appearing in the last integral. Since only the component
of $\mathbf{u}$ normal to the surface $\partial \mathfrak{b}\left( t\right) $%
\ is independent of the parametrization of $\partial \mathfrak{b}\left(
t\right) $, a natural invariance requirement with respect to such a
parametrization implies that the vector $\left( \mathbb{P+}\mathbf{F}^{T}%
\mathbf{T+}\nabla \mathbf{\nu }^{T}\underline{\ast }\mathcal{S}\right) 
\mathbf{n}$ must be purely normal to $\partial \mathfrak{b}$; then there
exists a scalar $\varpi $ such that $\mathbb{P+}\mathbf{F}^{T}\mathbf{T+}%
\nabla \mathbf{\nu }^{T}\underline{\ast }\mathcal{S}=\varpi \mathbf{I}$. By
substituting $\varpi \mathbf{I}$ within the previous inequality, we find the
term $\left( \psi -\varpi \right) \left( \mathbf{u\cdot n}\right) $.
However, since (44) is assumed to hold for any choice of the velocity
fields, we get $\psi =\varpi $, then $\mathbb{P}=\psi \mathbf{I-F}^{T}%
\mathbf{T-}\nabla \mathbf{\nu }^{T}\underline{\ast }\mathcal{S}$. Notice
that, in absence of prominent effects of the material substructure described
by $\mathbf{\nu }$, the second order tensor $\mathbb{P}$ reduces to the
well-known Eshelby tensor $\psi \mathbf{I-F}^{T}\mathbf{T}$.

We restrict our analysis to the non-homogeneous purely non-linear elastic
case and assume constitutive expressions of the form $\mathbf{T=\hat{T}}%
\left( \mathbf{X},\mathbf{F,\nu ,}\nabla \mathbf{\nu }\right) $ for the
Piola-Kirchhoff stress, $\mathbf{z=\hat{z}}\left( \mathbf{X},\mathbf{F,\nu ,}%
\nabla \mathbf{\nu }\right) $ for the self-force and $\mathcal{S}=\overset{%
\frown }{\mathcal{S}}\left( \mathbf{X},\mathbf{F,\nu ,}\nabla \mathbf{\nu }%
\right) $ for the microstress. If we take the free energy as $\psi =\hat{\psi%
}\left( \mathbf{X},\mathbf{F,\nu ,}\nabla \mathbf{\nu }\right) $, and assume
that it admits partial derivatives with respect to its entries, with the use
of previous results, the mechanical dissipation inequality reduces to%
\begin{equation}
\int_{\mathfrak{b}\left( t\right) }\left( \left( \partial _{\mathbf{F}}\psi -%
\mathbf{T}\right) \cdot \mathbf{\dot{F}+}\left( \partial _{\mathbf{\nu }%
}\psi -\mathbf{z}\right) \cdot \mathbf{\dot{\nu}}\right) +\int_{\mathfrak{b}%
\left( t\right) }\left( \left( \rho \partial _{\nabla \mathbf{\nu }}\psi -%
\mathcal{S}\right) \cdot \nabla \mathbf{\dot{\nu}}\right) \leq 0,
\end{equation}%
where $\partial _{y}\psi $ means partial derivative of $\psi $\ with respect
to the argument $y$. Its validity for any choice of the rates implies%
\begin{equation}
\mathbf{T}=\partial _{\mathbf{F}}\psi \text{ \ \ \ \ ; \ \ \ \ }\mathbf{z}%
=\partial _{\mathbf{\nu }}\psi \text{ \ \ \ \ ; \ \ \ \ }\mathcal{S}%
=\partial _{\nabla \mathbf{\nu }}\psi .
\end{equation}%
As a consequence, taking into account the explicit expression of $\mathbb{P}$%
, from (35) we get $\mathbf{g=-}\partial _{\mathbf{X}}\psi $ and $\mathbf{%
e=-F}^{T}\mathbf{b-}\left( \nabla \mathbf{\nu }^{T}\right) \mathbf{\beta }$.

Special expressions of $\psi $ are of Ginzburg-Landau type. In most cases,
in fact, it seems to be natural to assume $\psi =\check{\psi}\left( \mathbf{X%
},\mathbf{F,\nu }\right) +\frac{1}{2}a\left( \mathbf{X}\right) \left\Vert
\nabla \mathbf{\nu }\right\Vert ^{2}$. In particular, if we have $\check{\psi%
}\left( \mathbf{X},\mathbf{F,\nu }\right) =\check{\psi}_{1}\left( \mathbf{X},%
\mathbf{F}\right) +\check{\psi}_{2}\left( \mathbf{\nu }\right) $, $a$
constant and $\check{\psi}_{2}\left( \mathbf{\nu }\right) $\ a
coarse-grained (perhaps multiwell) energy, the balance of substructural
interactions (14) coincides with the well known Ginzburg-Landau equation.

Viscous effects may occur at the gross scale and at the substructural level.
In this case, the measures of interaction, $\mathbf{T}$, $\mathbf{z}$, $%
\mathcal{S}$ may depend on the rates of the fields and their gradients. We
assume, as a prototype example, that only the self-force $\mathbf{z}$
depends on the sole rate $\mathbf{\dot{\nu}}$. The self-force $\mathbf{z}$
may be thus decomposed into its viscous ($v$) and non-viscous ($nv$) parts,
namely $\mathbf{z=z}^{v}+\mathbf{z}^{nv}$, with $\mathbf{z}^{nv}=\mathbf{z}%
^{nv}\left( \mathbf{X},\mathbf{F,\nu ,}\nabla \mathbf{\nu }\right) $ and $%
\mathbf{z}^{v}=\mathbf{z}^{v}\left( \mathbf{X},\mathbf{F,\nu ,}\nabla 
\mathbf{\nu ;\dot{\nu}}\right) $, with $\mathbf{z}^{v}\cdot \mathbf{\dot{\nu}%
\geq }0$ for any choice of $\mathbf{\dot{\nu}}$, which implies $\mathbf{z}%
^{v}=\lambda \left( \mathbf{X},\mathbf{F,\nu ,}\nabla \mathbf{\nu }\right) 
\mathbf{\dot{\nu}}$, with $\lambda $ some positive scalar function and $%
\mathbf{z}^{nv}$\ satisfying (46b).

\subsection{The mechanical dissipation inequality along the sides of the
crack. Consequences.}

For a part $\mathfrak{b}_{\mathcal{C}}\left( t\right) $\ crossing the crack
away the tip, we consider an additional surface free energy density $\phi
\left( \mathbf{X}\right) $ along the margins of the crack; it is continuous
up to the tip where it is indicated with $\phi _{tip}$. As a consequence, to
the bulk terms, the ones in (44), we must add the piece%
\begin{equation}
\int_{\mathfrak{b}_{\mathcal{C}}\left( t\right) \cap \mathcal{C}\left(
t\right) }\dot{\phi}+\int_{\left( \partial \mathfrak{b}_{\mathcal{C}}\left(
t\right) \right) \cap \mathcal{C}\left( t\right) }\phi \left( \mathfrak{m}%
\cdot \mathbf{u}\right) -\int_{\left( \partial \mathfrak{b}_{\mathcal{C}%
}\left( t\right) \right) \cap \mathcal{C}\left( t\right) }\bar{\sigma}\left( 
\mathfrak{m}\cdot \mathbf{u}\right) .
\end{equation}%
We may then reduce the resulting inequality by shrinking $\mathfrak{b}_{%
\mathcal{C}}\left( t\right) $\ to $\mathcal{C}\left( t\right) $, and taking
the limit uniformly in time. In this case we get%
\begin{equation}
\int_{\left( \partial \mathfrak{b}_{\mathcal{C}}\left( t\right) \right) \cap 
\mathcal{C}\left( t\right) }\left( \phi -\bar{\sigma}\right) \left( 
\mathfrak{m}\cdot \mathbf{u}\right) -\int_{\mathfrak{b}_{\mathcal{C}}\left(
t\right) \cap \mathcal{C}}\left( \left[ \mathbf{Tm\cdot \dot{x}}\right] +%
\left[ \mathcal{S}\mathbf{m\cdot \dot{\nu}}\right] \right) \leq 0.
\end{equation}%
The validity of such an inequality for any choice of velocity fields implies 
$\phi =\bar{\sigma}$ and the local dissipation inequality $\left[ \mathbf{%
Tm\cdot \dot{x}}\right] +\left[ \mathcal{S}\mathbf{m\cdot \dot{\nu}}\right]
\leq 0$.

\subsection{The mechanical dissipation inequality at the tip of the crack.
Consequences.}

Let $\mathfrak{b}_{R}^{\ast }\left( t\right) $\ a `curved cylinder' wrapped
around the tip as used in previous sections; its boundary intersects the tip
in two points $\mathbf{Z}\left( \mathfrak{s}_{1}\left( t\right) \right) $
and $\mathbf{Z}\left( \mathfrak{s}_{2}\left( t\right) \right) $. In writing
the mechanical dissipation inequality on $\mathfrak{b}_{R}^{\ast }\left(
t\right) $, we consider, in addition to bulk and surface energies, a line
energy density $\zeta $; moreover, since the tip moves, we must account also
for the power of \ $\lambda _{tip}\mathsf{t}$ and $\mathbf{e}_{tip}$ ($%
\mathbf{g}_{tip}$ is excluded because it is internal). Consequently, to bulk
and surface contributions we add the term 
\begin{equation}
\frac{d}{dt}\left( \int_{\mathfrak{s}_{1}\left( t\right) }^{\mathfrak{s}%
_{2}\left( t\right) }\zeta \right) -\left( \lambda _{tip}\mathsf{t}\cdot 
\mathbf{v}_{tip}\left\vert _{\mathfrak{s=s}_{2}}\right. -\lambda _{tip}%
\mathsf{t}\cdot \mathbf{v}_{tip}\left\vert _{\mathfrak{s=s}_{1}}\right.
\right) .
\end{equation}%
By shrinking $\mathfrak{b}_{R}^{\ast }\left( t\right) $\ up to the tip,
taking the limit uniformly in time and making use of the line balance (39),
we get%
\begin{equation*}
\left( \zeta -\lambda _{tip}\right) \left( \mathsf{t}\cdot \mathbf{v}%
_{tip}\left\vert _{\mathfrak{s=s}_{2}}\right. -\mathsf{t}\cdot \mathbf{v}%
_{tip}\left\vert _{\mathfrak{s=s}_{1}}\right. \right) +
\end{equation*}%
\begin{equation}
+\int_{\mathfrak{s}_{1}\left( t\right) }^{\mathfrak{s}_{1}\left( t\right)
}\zeta \mathfrak{h\cdot }\mathbf{v}_{tip}+\int_{\mathfrak{s}_{1}\left(
t\right) }^{\mathfrak{s}_{2}\left( t\right) }\mathbf{v}_{tip}\cdot \mathbf{g}%
_{tip}\leq 0
\end{equation}%
Since the resulting inequality must be valid for any choice of the velocity
fields, then of $\mathbf{v}_{tip}$, we obtain $\zeta =\lambda _{tip}$ and $%
\mathbf{g}_{tip}\cdot \mathbf{v}_{tip}\leq 0$, which reduces to $V\mathbf{g}%
_{tip}\cdot \mathsf{n}\leq 0$ since $\mathbf{v}_{tip}=V\mathsf{n}$, implying
then that the component of $\mathbf{g}_{tip}$ along the motion of $\mathcal{J%
}$, namely $g_{tip}=\mathbf{g}_{tip}\cdot \mathsf{n}$, must have a structure
of the type $g_{tip}=\mathfrak{g}_{tip}V$, where $\mathfrak{g}_{tip}$\ is a
negative `diffusion' coefficient that must be assigned constitutively.

\section{Driving the tip of the crack}

\subsection{The driving force}

The explicit expressions of $\mathbb{P}$, $\mathbf{g}$, $\bar{\sigma}$, $%
\lambda _{tip}$ allow us to write the tip balance (39) as 
\begin{equation}
-\phi _{tip}\mathsf{n}-\lambda _{tip}\mathfrak{h+}\int_{tip}\left( \rho
\left( \frac{1}{2}\left\Vert \mathbf{\dot{x}}\right\Vert ^{2}+k\left( 
\mathbf{\nu },\mathbf{\dot{\nu}}\right) \right) \mathbf{I-}\mathbb{P}\right) 
\mathbf{n}=-\mathbf{g}_{tip}.
\end{equation}%
We indicate with $\mathbf{j}$\ the vector 
\begin{equation}
\mathbf{j=}\int_{tip}\left( \rho \left( \frac{1}{2}\left\Vert \mathbf{\dot{x}%
}\right\Vert ^{2}+k\left( \mathbf{\nu },\mathbf{\dot{\nu}}\right) \right) 
\mathbf{I-}\mathbb{P}\right) \mathbf{n}
\end{equation}%
It represents the \emph{tip traction} exerted by the bulk material on an
infinitesimal neighborhood around the tip. Let $\mathsf{n}\left( \mathfrak{s}%
\right) $\ be the \emph{direction of propagation of the crack} at the point $%
\mathbf{Z}\left( \mathfrak{s}\right) $ of the tip, the component of (51)
along $\mathsf{n}\left( \mathfrak{s}\right) $\ is given by 
\begin{equation}
-\phi _{tip}-\lambda _{tip}\mathfrak{K}+\mathsf{n}\cdot \mathbf{j}=-%
\mathfrak{g}_{tip}V.
\end{equation}%
By indicating with $\mathsf{J}$\ (the \textsf{J}-integral) the product $%
\mathsf{n}\cdot \mathbf{j}$, we interpret the difference 
\begin{equation}
\mathfrak{f}=\mathsf{J}-\phi _{tip}-\lambda _{tip}\mathfrak{K}
\end{equation}%
as the \emph{force driving the tip of the crack}; it accounts directly for
the influence of the material substructure. Since $\mathbf{g}_{tip}\cdot 
\mathbf{v}_{tip}\leq 0$, we get $\mathfrak{f}V\geq 0$ (we remind that $%
\mathbf{v}_{tip}$ is of the form $V\mathsf{n}$). When the crack grows, i.e.,
when $V>0$, the driving force must be non-negative, i.e., $\mathfrak{f}\geq 0
$.

\subsection{Dynamic energy release rate at the tip.}

The energy release rate at the tip is given by the power $\mathfrak{f}V$\
developed by the driving force along the normal motion of the crack tip.
Previous results allow us to express the product $\mathfrak{f}V$\ in terms
of the power of standard and substructural interactions and of the free
energy, and we obtain 
\begin{equation*}
\int_{tip}\left( \rho \left( \psi +\frac{1}{2}\left\Vert \mathbf{\dot{x}}%
\right\Vert ^{2}+k\left( \mathbf{\nu },\mathbf{\dot{\nu}}\right) \right) V+%
\mathbf{Tn\cdot \dot{x}}+\mathcal{S}\mathbf{n\cdot \dot{\nu}}\right) -
\end{equation*}%
\begin{equation}
-\phi _{tip}V-\lambda _{tip}\mathfrak{K}V=\mathfrak{f}V,
\end{equation}%
which represents \emph{the balance of energy at the tip}.

\subsection{Quasi-static extended \textsf{J}-integral and its path
independence}

When inertial effects are negligible, the evolution of the crack is
\textquotedblleft quasi static\textquotedblright . The $\mathsf{J}$-integral
($\mathsf{J}=\mathsf{n}\cdot \mathbf{j}$) reduces to its quasi static
counterpart $\mathsf{J}_{qs}$: 
\begin{equation}
\mathsf{J}_{qs}=\mathsf{n}\cdot \int_{tip}\mathbb{P}\mathbf{n.}
\end{equation}

It reduces to the standard $\mathsf{J}$-integral\ given by $\mathsf{n}\cdot
\int_{tip}\left( \psi \mathbf{I-F}^{T}\mathbf{T}\right) \mathbf{n}$ when the
substructure is absent or its gross effects are negligible.

\emph{If the material is homogeneous, }$\mathcal{C}$\emph{\ is planar (i.e.
the crack is straight), the crack has the margins free of standard and
substructural tractions (in the sense that }$\mathbf{T}^{\pm }\mathbf{m=0}$ 
\emph{and} $\mathcal{S}^{\pm }\mathbf{m=0}$\emph{),} $\mathsf{J}_{qs}$\ 
\emph{is path-independent}.

The hypotheses of homogeneity of the material and absence of inertial
effects imply $\mathbf{g=0}$ and $\mathbf{e=0}$. Then, the bulk balance (35)
reduces to $Div\mathbb{P}=0$. Moreover, since the margins of the crack are
free of standard and substructural tractions, we get $\left[ \mathbb{P}%
\right] \mathbf{m=}\left[ \psi \right] \mathbf{m}$.

With these premises, we take in $\mathcal{B}$ an arbitrary `curved cylinder' 
$\mathfrak{b}_{R}^{\ast }$\ wrapped around the tip where we have%
\begin{equation}
\mathsf{n}\cdot \int_{\partial \mathfrak{b}_{R}^{\ast }\left( t\right) }%
\mathbb{P}\mathbf{n}=\mathsf{n}\cdot \int_{\mathfrak{s}_{1}}^{\mathfrak{s}%
_{2}}\int_{\partial D_{R}}\mathbb{P}\mathbf{n.}
\end{equation}%
$D_{R}$ is arbitrary, then we need only to evaluate the difference 
\begin{equation}
\mathsf{n}\left( \mathfrak{s}\right) \cdot \int_{\partial D_{R}}\mathbb{P}%
\mathbf{n-}\mathsf{n}\left( \mathfrak{s}\right) \cdot \int_{tip}\mathbb{P}%
\mathbf{n,}
\end{equation}%
that is equal to $\mathsf{n\cdot }\int_{D_{R}\cap \mathcal{C}}\left[ \psi %
\right] \mathbf{m}$, which vanishes because the crack is straight, i.e. $%
\mathbf{m\perp }\mathsf{n}$. The path-independence of $\mathsf{J}_{qs}$
follows.

\section{The energy dissipated into a process zone of finite size around the
crack tip}

When a crack propagates, a material part $Pz$ around the tip becomes highly
unstable, in certain sense `fragmented' (Aoki, Kishimoto \& Sakata, 1981;
1984). Usually, $Pz$ is called \emph{process zone}. In brittle fracture, the
process zone may be considered practically coincident with the tip, while in
ductile fracture $Pz$ has finite size. For the latter case we obtain new
path-integrals which allow us to evaluate the energy dissipated during the
evolution of the crack.

We assume that $Pz$ can be approximated reasonably by a `curved cylinder' $P$%
\ wrapped around $\mathcal{J}$ (basically, $P$ has geometrical properties
analogous to $\mathfrak{b}_{R}^{\ast }$\ used in previous sections); the
intersection of $P$ with the plane orthogonal to the tangent $\mathsf{t}%
\left( \mathfrak{s}\right) $\ of $\mathcal{J}$ at $\mathfrak{s}\in \left[ 0,%
\mathfrak{\bar{s}}\right] $ is a disc $P^{\pi }$ with the centre on $%
\mathcal{J}$ (being the centre the sole intersection of $\mathcal{J}$ with $P
$).\ The approximation of $Pz$\ with $P$ is rather rough but it does not
influence the basic structure of the results obtained in the present
section. During the evolution of the crack, $P$ varies in time and is $%
P\left( t\right) $. The boundary $\partial P$ is a surface (with outward
unit normal indicated with $\mathbf{n}$) parametrized by $\upsilon _{1}$, $%
\upsilon _{2}$\ and points $\mathbf{\bar{X}}\left( \upsilon _{1},\upsilon
_{2},t\right) \in \partial P\left( t\right) $ have an intrinsic velocity $%
\mathbf{\hat{u}}$ given by $\mathbf{\hat{u}=\partial }_{t}\mathbf{\bar{X}}%
\left( \upsilon _{1},\upsilon _{2},t\right) $. Consequently, rates following 
$\partial P\left( t\right) $\ are $\mathbf{x}^{\triangleright }=\mathbf{\dot{%
x}+F\hat{u}}$, and $\mathbf{\nu }^{\triangleright }=\mathbf{\dot{\nu}+}%
\left( \nabla \mathbf{\nu }\right) \mathbf{\hat{u}}$.

The velocity $\mathbf{\hat{u}}$\ at the boundary $\mathfrak{\partial }\bar{P}
$ is decomposed as%
\begin{equation}
\mathbf{\hat{u}=\hat{u}}_{tr}\left( t\right) +\mathbf{\dot{q}}\left(
t\right) \mathbf{\times }\left( \mathbf{X-X}_{0}\right) +\alpha \left(
t\right) \left( \mathbf{X-X}_{0}\right) +\mathbf{\hat{u}}_{d}\left( \mathbf{X%
},t\right) ,
\end{equation}%
$\mathbf{\hat{u}}_{tr}\left( t\right) $ denotes the component of rigid
translation; $\mathbf{\dot{q}}\left( t\right) \mathbf{\times }\left( \mathbf{%
X-X}_{0}\right) $ the rotational component, with $\mathbf{X}_{0}$ an
arbitrary fixed point (note that the presence of $\left( \mathbf{X-X}%
_{0}\right) $ instead of $\left( \mathbf{x-x}_{0}\right) $ is due to the
circumstance that $\mathbf{\hat{u}}$\ is a material velocity in $\mathcal{B}$%
); $\alpha \left( t\right) \left( \mathbf{X-X}_{0}\right) $ the velocity
associated with the self-similar expansion of $P$; $\mathbf{\hat{u}}%
_{d}\left( \mathbf{X},t\right) $ the component of the velocity due to the
distortion.

Let now\ $\mathfrak{b}_{R}^{\ast }$\ be another `curved cylinder' wrapped
around the tip, \emph{fixed in time} and containing a piece $\bar{P}$ of the
process zone. We analyze the behavior of $\bar{P}\left( t\right) $ in a time
interval in which $\partial \bar{P}\left( t\right) $ does not intersect $%
\partial \mathfrak{b}_{R}^{\ast }$.

With reference to the situation described before, we indicate with $\hat{\Phi%
}\left( \bar{P}\right) $ the \emph{rate of energy dissipated in} $\bar{P}$\
during the evolution of the crack and assume that \emph{all the dissipation
is concentrated in }$P$ \emph{during the evolution of the crack}. Such an
assumption implies that 
\begin{equation}
\frac{d}{dt}\int_{\mathfrak{b}_{R}^{\ast }\backslash \bar{P}\left( t\right)
}\psi =\int_{\mathfrak{b}_{R}^{\ast }\backslash \bar{P}\left( t\right)
}\left( \mathbf{T}\cdot \mathbf{\dot{F}+z}\cdot \mathbf{\dot{\nu}+}\mathcal{S%
}\cdot \nabla \mathbf{\dot{\nu}}\right) .
\end{equation}%
In other words, the mechanical dissipation inequality in $\mathfrak{b}%
_{R}^{\ast }\backslash P\left( t\right) $\ reduces to an equation because no
dissipation mechanism occur outside $\bar{P}$.

With these premises, the balance of the energy takes the form%
\begin{equation*}
\frac{d}{dt}\int_{\mathfrak{b}_{R}^{\ast }\backslash \bar{P}\left( t\right)
}\rho \left( \frac{1}{2}\left\Vert \mathbf{\dot{x}}\right\Vert ^{2}+k\left( 
\mathbf{\nu },\mathbf{\dot{\nu}}\right) \right) +\int_{\mathfrak{\partial }%
\bar{P}\left( t\right) }\rho \left( \frac{1}{2}\left\Vert \mathbf{\dot{x}}%
\right\Vert ^{2}+k\left( \mathbf{\nu },\mathbf{\dot{\nu}}\right) \right)
\left( \mathbf{\hat{u}\cdot n}\right) +
\end{equation*}%
\begin{equation*}
+\frac{d}{dt}\int_{\mathfrak{b}_{R}^{\ast }\backslash \bar{P}\left( t\right)
}\psi +\int_{\mathfrak{\partial }\bar{P}\left( t\right) }\psi \left( \mathbf{%
\hat{u}\cdot n}\right) -\hat{\Phi}\left( \bar{P}\right) =
\end{equation*}%
\begin{equation}
=\int_{\mathfrak{b}_{R}^{\ast }\backslash \bar{P}\left( t\right) }\left( 
\mathbf{b}^{ni}\mathbf{\cdot \dot{x}+\beta }^{ni}\mathbf{\cdot \dot{\nu}}%
\right) +\int_{\partial \mathfrak{b}_{R}^{\ast }}\left( \mathbf{Tn\cdot \dot{%
x}+}\mathcal{S}\mathbf{n\cdot \dot{\nu}}\right) .
\end{equation}

By making use of the weak form over $\mathfrak{b}_{R}^{\ast }\backslash \bar{%
P}\left( t\right) $\ of the balances of standard and substructural
interactions (11) and (14), and of the integral identity%
\begin{equation*}
\int_{\partial \mathfrak{b}_{R}^{\ast }}\left( \mathbf{Tn\cdot \dot{x}+}%
\mathcal{S}\mathbf{n\cdot \dot{\nu}}\right) =\int_{\mathfrak{\partial }\bar{P%
}\left( t\right) }\left( \mathbf{Tn}\cdot \mathbf{x}^{\triangleright }+%
\mathcal{S}\mathbf{n}\cdot \mathbf{\nu }^{\triangleright }\right) +
\end{equation*}%
\begin{equation}
+\int_{\mathfrak{b}_{R}^{\ast }\backslash \bar{P}\left( t\right) }\left(
Div\left( \mathbf{\dot{x}T}\right) +Div\left( \mathbf{\dot{\nu}}S\right)
\right) ,
\end{equation}%
from the arbitrariness of the piece of $\mathcal{J}$ considered, we obtain 
\begin{equation}
\Phi \left( P^{\pi }\right) =\mathbf{\hat{u}}_{tr}\left( t\right) \cdot 
\mathbf{j}\left( P^{\pi }\right) +\mathbf{\dot{q}}\left( t\right) \mathbf{%
\cdot L}+\alpha \left( t\right) M+I,
\end{equation}%
where 
\begin{equation}
\mathbf{j}\left( P^{\pi }\right) \mathbf{=}\int_{\mathfrak{\partial }P^{\pi
}\left( t\right) }\left( \rho \left( \frac{1}{2}\left\Vert \mathbf{\dot{x}}%
\right\Vert ^{2}+k\left( \mathbf{\nu },\mathbf{\dot{\nu}}\right) \right) 
\mathbf{I-}\mathbb{P}\right) \mathbf{n}
\end{equation}%
\begin{equation}
\mathbf{L=}\int_{\mathfrak{\partial }P^{\pi }\left( t\right) }\left( \mathbf{%
X-X}_{0}\right) \times \left( \rho \left( \frac{1}{2}\left\Vert \mathbf{\dot{%
x}}\right\Vert ^{2}+k\left( \mathbf{\nu },\mathbf{\dot{\nu}}\right) \right) 
\mathbf{I-}\mathbb{P}\right) \mathbf{n}
\end{equation}%
\begin{equation}
M\mathbf{=}\int_{\mathfrak{\partial }P^{\pi }\left( t\right) }\left( \rho
\left( \frac{1}{2}\left\Vert \mathbf{\dot{x}}\right\Vert ^{2}+k\left( 
\mathbf{\nu },\mathbf{\dot{\nu}}\right) \right) \mathbf{I-}\mathbb{P}\right) 
\mathbf{n\cdot }\left( \mathbf{X-X}_{0}\right) 
\end{equation}%
\begin{equation}
I\mathbf{=}\int_{\mathfrak{\partial }P^{\pi }\left( t\right) }\left( \rho
\left( \frac{1}{2}\left\Vert \mathbf{\dot{x}}\right\Vert ^{2}+k\left( 
\mathbf{\nu },\mathbf{\dot{\nu}}\right) \right) \mathbf{I-}\mathbb{P}\right) 
\mathbf{n\cdot \hat{u}}_{d}-\int_{\mathfrak{\partial }P^{\pi }\left(
t\right) }\left( \mathbf{Tn}\cdot \mathbf{\dot{x}}+\mathcal{S}\mathbf{n}%
\cdot \mathbf{\dot{\nu}}\right) .
\end{equation}

\section{Special cases}

The theory discussed in previous sections allows us to describe the behavior
of cracks in several cases of complex materials. Two essential ingredients
are necessary to apply the results: (i) the choice of an order parameter $%
\mathbf{\nu }$ (hence of $\mathcal{M}$) describing appropriately the
material substructure; (ii) an explicit expression of the free energy. In
what follows, we indicate two possible spheres of application: ferroelectric
solids and solids exhibiting strain gradient effects.

\subsection{Cracks in ferroelectrics}

\subsubsection{Preliminary remarks}

The predictions of Griffith's theory fall for cracks propagating in
ferroelectric solids subjected to the action of external electric fields:
there is a discrepancy between the driving force predicted theoretically and
the experimental data. Basically, we regard this discrepancy as due to the
circumstance that Griffith's theory does not account for the substructural
interactions associated with the occurrence of spontaneous polarization that
can be induced by strain, variation of temperature and applied electric
fields. The polarization is indicated with $\mathbf{p}$ and we take as \emph{%
order parameter}\ the vector $\mathsf{p}=\rho ^{-1}\mathbf{p}$ such that $%
0\leq \left\vert \mathsf{p}\right\vert \leq p_{m}$, with $p_{m}$ a material
constant. Then $\mathcal{M}$ is the ball of radius $p_{m}$ in $\mathbb{R}%
^{3} $. We also consider the body as subjected to an external electric field 
$\mathfrak{E}$.

Balance equations are formally identical to (11), (13) and (14). In this
case, $\mathcal{A}$ is the second order tensor $\mathsf{p}\times $ (in
components $\left( \mathsf{p}\times \right) _{ij}=\mathsf{e}_{ijk}\mathsf{p}%
_{k}$, with $\mathsf{e}_{ijk}$ the alternating symbol). As a consequence,
the relation (5) becomes $\mathsf{\dot{p}}^{\ast }=\mathsf{\dot{p}}+\mathsf{p%
}\times \mathbf{\dot{q}}$. Here, the microstress $\mathcal{S}$ accounts for
interactions between neighboring crystals with different polarizations; $%
\mathbf{z}$ measures self-interactions within each polarized crystal. To
include the effects of the applied electric field in the balance equations
we assume that the bulk interactions $\mathbf{b}$ and $\mathbf{\beta }$ and
the boundary `tractions' $\mathbf{t=Tn}$ and $\mathbf{\tau =}\mathcal{S}%
\mathbf{n}$ can be decomposed additively in electromechanical parts $\left(
em\right) $ and purely electric parts $\left( el\right) $: 
\begin{equation}
\mathbf{b=b}_{em}\mathbf{+b}_{el}\text{ \ \ \ \ ; \ \ \ \ }\mathbf{\beta
=\beta }_{em}\mathbf{+\beta }_{el};
\end{equation}%
\begin{equation}
\mathbf{t=t}_{em}\mathbf{+t}_{el}\text{ \ \ \ \ ; \ \ \ \ }\mathbf{\tau
=\tau }_{em}\mathbf{+\tau }_{el}.
\end{equation}

For any arbitrary part $\mathfrak{b}$ of $\mathcal{B}$,\ the purely electric
parts are characterized by the balance%
\begin{equation}
\frac{d}{dt}D\left( \mathfrak{b}\right) +\int_{\mathfrak{b}}\left( \mathbf{b}%
_{el}\mathbf{\cdot \dot{x}+\mathbf{\beta }}_{el}\mathbf{\cdot }\mathsf{\dot{p%
}}\right) +\int_{\partial \mathfrak{b}}\left( \mathbf{t}_{el}\mathbf{\cdot 
\dot{x}+\mathbf{\tau }}_{el}\mathbf{\cdot }\mathsf{\dot{p}}\right) =0,
\end{equation}%
where $D\left( \mathfrak{b}\right) =\frac{1}{2}\int_{\mathfrak{b}}\left\vert 
\mathfrak{E}\right\vert =-\frac{1}{2}\int_{\mathfrak{b}}\rho \mathfrak{%
E\cdot }\mathsf{p}$ is the \emph{electric energy} of $\mathfrak{b}$.

A theorem of Tiersten (1964) give us the explicit expression of the rate of $%
D$ in the current configuration. We indicate the material version of
Tiersten's formula by pulling it back in the reference configuration:%
\begin{equation}
\frac{d}{dt}D\left( \mathfrak{b}\right) =-\int_{\mathfrak{b}}\rho \left( grad%
\mathfrak{E}\right) \mathsf{p}\cdot \mathbf{\dot{x}-}\int_{\partial 
\mathfrak{b}}\frac{1}{2}\left( \det \mathbf{F}\right) p_{n}^{2}\mathbf{F}%
^{-T}\mathbf{n\cdot \dot{x}-}\int_{\mathfrak{b}}\rho \mathfrak{E}\cdot 
\mathsf{\dot{p}},
\end{equation}%
where $p_{n}$ is the normal component of $\mathsf{p}$, $grad$ indicates the
gradient with respect to $\mathbf{x}$. It follows that 
\begin{equation}
\mathbf{b}_{el}=\rho \left( grad\mathfrak{E}\right) \mathsf{p}\text{ \ \ \ \
; \ \ \ \ \ }\mathbf{\mathbf{\beta }}_{el}=\rho \mathfrak{E};
\end{equation}%
\begin{equation}
\mathbf{t}_{el}=\frac{1}{2}\left( \det \mathbf{F}\right) p_{n}^{2}\mathbf{F}%
^{-T}\mathbf{n}\text{ \ \ \ \ ; \ \ \ \ \ }\mathbf{\mathbf{\tau }}_{el}=0.
\end{equation}%
Consequently, the balance equations (11) and (14) become 
\begin{equation}
\mathbf{b}_{em}+Div\mathbf{T}+\rho \left( grad\mathfrak{E}\right) \mathsf{p}%
=0,
\end{equation}%
\begin{equation}
\mathbf{\beta }_{em}-\mathbf{z+}Div\mathcal{S}+\rho \mathfrak{E}=0,
\end{equation}%
where $\mathbf{b}_{em}$\ and $\mathbf{\beta }_{em}$ include the inertial
terms as ever (see also Dav\`{\i}, 2001).

\subsubsection{The driving force in ferroelectrics}

Taking in mind the identification of $\mathbf{\nu }$\ with $\mathsf{p}$, all
the general results on cracks presented before apply and for non-linear
elastic ferroelectrics the quasi-static \textsf{J}-integral is given by 
\begin{equation}
\mathsf{J}_{qs}=\mathsf{n}\cdot \int_{tip}\left( \psi \mathbf{I-F}%
^{T}\partial _{\mathbf{F}}\psi \mathbf{-}\nabla \mathsf{p}^{T}\partial
_{\nabla \mathsf{p}}\psi \right) \mathbf{n}
\end{equation}%
An expression of this type has been used in (Wei \& Hutchinson, 1997) for
the special case of infinitesimal strains and fits reasonably the
experimental data about the driving force.

\subsection{Cracks in materials with strain gradient effects}

Size effects are well recognized in the behavior of crystalline solids even
during phase transitions. To describe these experimental evidences, models
involving the second gradient of deformation have been considered (Dunn \&
Serrin, 1985). Strain gradient effects are induced by a \emph{latent}
substructure. Following the thermodynamically consistent theory of Capriz
(1985), latence is induced here by an internal constraint obtained by
identifying $\mathbf{\nu }$ with $\mathbf{F}$, in absence of external
actions on the substructure ($\mathbf{\beta =0}$). As a consequence, if one
assumes constitutive equations of the type $\mathbf{T=\hat{T}}\left( \mathbf{%
F},\nabla \mathbf{F}\right) $, $\mathcal{S}=\mathcal{\breve{S}}\left( 
\mathbf{F},\nabla \mathbf{F}\right) $ and $\psi =\hat{\psi}\left( \mathbf{F}%
,\nabla \mathbf{F}\right) $, it follows that $\mathbf{T}=\partial _{\mathbf{F%
}}\psi $, $\mathcal{S}=\partial _{\nabla \mathbf{F}}\psi $ and the balances
of standard and substructural interactions merge one into the other and
reduce to the sole balance%
\begin{equation}
Div\left( \mathbf{T-}Div\mathcal{S}\right) +\mathbf{b}=\mathbf{0.}
\end{equation}

The quasi static \textsf{J}-integral \textsf{J}$_{qs}$ takes the special
form 
\begin{equation}
\mathsf{J}_{qs}=\mathsf{n}\cdot \int_{tip}\left( \psi \mathbf{I-F}%
^{T}\partial _{\mathbf{F}}\psi \mathbf{-}\nabla \mathbf{F}^{T}:\partial
_{\nabla \mathbf{F}}\psi \right) \mathbf{n;}
\end{equation}%
its reduced version in infinitesimal deformations has been used in (Fulton
\& Gao, 2001) to calculate the driving force in crystalline materials that
fail by decohesion at atomic scale: the results fit reasonably experimental
data.

\section{Discussion}

We have presented a fully non-linear three-dimensional description of the
crack growth under large strains in complex bodies suffering a prominent
influence of the material substructure on the gross behavior. We have used
`abstract' order parameter fields as coarse-grained descriptors of the
morphology of the substructure. From one hand such a point of view allows us
to unify some existing preliminary tentative to modify the standard theory
of fracture in special cases of complex bodies, while, on the other hand, it
furnishes a general tool able to provide directly the expression of the
driving force in all cases of materials admitting Ginzburg-Landau-like
energies. The present paper extends and renders more perspicuous preliminary
two-dimensional results presented in a section of the article (Mariano,
2001).

\ \ \ \ \ \ \ \ \ \ \ \ \ 

\textbf{Acknowledgment}. I wish to thank for valuable discussions Davide
Bernardini and Furio L. Stazi. The support of the Italian National Group of
Mathematical Physics (GNFM) is gratefully acknowledged.

\section{References}

\begin{description}
\item Abeyaratne, R., Knowles, J. K. 1990 On the diving traction acting on a
surface of strain discontinuity in a continuum. \emph{J. Mech. Phys. Solids} 
\textbf{38}, 345-360.

\item Adda-Bedia, M., Arias, R., Ben Amar, M. \& Lund, F. 1999 Generalized
Griffith criterion for dynamic fracture and the stability of crack motion at
high velocities. \emph{Phys.l Rev. E} \textbf{60}, 2366-2376.

\item Atkinson, C. \& Eshelby, J. D. 1968 The flow of energy into the tip of
a moving crack. \emph{Int. J. Fracture} \textbf{4}, 3-8.

\item Aoki, S., Kishimoto, K. \& Sakata, M. 1981 Energy release rate in
elastic plastic fracture problems. ASME - \emph{J. Appl. Mech.} \textbf{48},
825-829.

\item Aoki, S., Kishimoto, K. \& Sakata, M. 1984 Energy flux into process
region in elastic plastic fracture problems. \emph{Engng Fracture Mech.} 
\textbf{19}, 827-836.

\item Barenblatt, G. I. 1972 The mathematical theory of equilibrium cracks
in brittle fracture. \emph{Adv. Appl. Mech.} \textbf{7}, 55-129.

\item Capriz, G. 1985 Continua with latent microstructure. \emph{Arch.
Rational Mech. Anal.} \textbf{90}, 43-56.

\item Capriz, G. 1989 \emph{Continua with microstructure}. Springer.

\item Dav\`{\i}, F. 2001 On domain switching in deformable ferroelectrics,
seen as continua with microstructure. \emph{Z. angew. Math. Phys. ZAMP} 
\textbf{52}, 966-989.

\item Dunn, J. E. \& Serrin, J. 1985 On the thermodynamics of interstitial
working. \emph{Arch. Rational Mech. Anal.} \textbf{88}, 95-133.

\item Dolbow, J., Mo\"{e}s, M. \& Belytschko, T. 2001 An extended finite
element method for modeling crack growth with frictional contact. \emph{%
Comp. Meth. Appl. Mech. Eng.} \textbf{190}, 6825-6846.

\item Epstein, M. 2002 The Eshelby tensor and the theory of continuous
distributions of inhomogeneities. \emph{Mech. Res. Comm.} \textbf{29},
501-506.

\item Eshelby, J. D. 1975 The elastic energy-momentum tensor. \emph{J.
Elasticity} \textbf{5}, 321-335.

\item Freund, L. B. 1972 Energy flux into the tip of an extending crack in
elastic solids. \emph{J. Elasticity} \textbf{2}, 321-349.

\item Freund, L. B. 1990 \emph{Dynamic fracture mechanics}. Cambridge
University Press.

\item Fulton, C. C. \& Gao, H. 2001 Microstructural modeling of
ferroelectric fracture. \emph{Acta Materialia} \textbf{49}, 2039-2054.

\item Griffith, A. A. 1920 The phenomenon of rupture and flow in solids. 
\emph{Phil. Trans. Roy. Soc. London Ser. A} \textbf{221}, 163-198.

\item Gurtin, M. E. 1995 The nature of configurational forces. \emph{Arch.
Rational Mech. Anal.} \textbf{131}, 67-100.

\item Gurtin, M. E. \& Shvartsman, M. M. 1997 Configurational forces and the
dynamics of planar cracks in three-dimensional bodies. \emph{J. Elasticity} 
\textbf{48}, 167-191.

\item Heino, P. \& Kaski, K. 1997 Dynamic fracture of disordered
viscoelastic solids. \emph{Phys. Rev. E} \textbf{56}, 4364-4370.

\item James, R. D. 2002 Configurational forces in magnetism with application
to the dynamics of a small scale ferromagnetic shape-memory cantilever. 
\emph{Cont. Mech. Therm.} \textbf{14}, 56-86.

\item Mariano, P. M. 2001 Multifield theories in mechanics of solids. \emph{%
Adv. Appl. Mech.} \textbf{38}, 1-93.

\item Mo\"{e}s, M. \& Belytschko, T. 2002 Extended finite element method for
cohesive crack growth. \emph{Eng. Frac. Mech.} \textbf{69}, 813-833.

\item Obrezanova, O., Movchan, A. B. \& Willis, J. R. 2003 Dynamic stability
of a propagating crack. \emph{J. Mech. Phys. Solids} \textbf{50}, 2637-2668.

\item Oleaga, G. E. 2003 On the dynamics of cracks in three dimensions. 
\emph{J. Mech. Phys. Solids }\textbf{51}, 169-185.

\item Rice, J. R. 1968 Mathematical analysis in the mechanics of fracture,
in \emph{Fracture} \textbf{2} (H. Liebowitz ed.) 191-311, Academic Press.

\item Segev, R. 1996 On smoothly growing bodies and the Eshelby tensor. 
\emph{Meccanica} \textbf{31}, 507-518.

\item \v{S}ilhav\'{y}, M. 1997 \emph{The mechanics and thermodynamics of
continuous media}. Springer.

\item Slepyan, L. I. 2002 \emph{Models and phenomena in fracture mechanics}.
Springer.

\item Tiersten, H. F. 1964 Coupled magnetomechanical equations for
magnetically saturated insulators. \emph{J. Math. Phys.} \textbf{5},
1298-1318.

\item Wei, Y. \& Hutchinson, J. W. 1997 Steady-state crack growth and work
of fracture for solids characterized by strain gradient plasticity. \emph{J.
Mech. Phys. Solids} \textbf{45}, 1253-1273.
\end{description}

\end{document}